\documentclass[twocolumn,11pt]{article}
\usepackage{graphicx}
\usepackage{amssymb}
\usepackage{color}

  \def\gtap{\mathrel{ \rlap{\raise 0.511ex \hbox{$>$}}{\lower 0.511ex
   \hbox{$\sim$}}}} 
\def\ltap{\mathrel{ \rlap{\raise 0.511ex
   \hbox{$<$}}{\lower 0.511ex \hbox{$\sim$}}}}

\newcommand{\tha}{\theta_{12}}
\newcommand{\thb}{\theta_{13}}
\newcommand{\thc}{\theta_{23}}
\newcommand{\dma}{\Delta m^2_{12}}

\newcommand{\dmc}{\Delta m^2_{23}}

\newcommand{\lt}{\leq}

\title{{\LARGE \bf The Neutrino Matrix}  \\}
\author{  Prepared by the Members of the APS Multidivisional Neutrino Study* \medskip \\}
\date{November, 2004 \\  }
\begin{document}
\ifx\href\undefined\else\hypersetup{linktocpage=true}\fi 
\begin{onecolumn}
\maketitle

\thispagestyle{empty}
\begin{centering}
\hspace{0.5in}
\mbox{
\frame{\hspace{0.1 in}
\begin{minipage}[]{5 in}
\vspace{0.1 in}
\noindent
{\bf matrix} {\it n:}\newline
1 : something within or from which something else originates, develops, \\
\phantom{aaa} or takes form \\
2 : the natural material in which something is embedded \\
3 : womb \\
4 : a rectangular array of mathematical elements 
\vspace{0.1 in}
\end{minipage}\hspace{0.1 in}
}}
\end{centering}
\vspace{3.8 in}
\begin{flushright}
*Please see Appendices A and B. \\
\vspace{0.2 in}
{\it Back Cover: Photomultipliers and Support Matrix \\ for the KamLAND Neutrino Detector.  \\ Courtesy KamLAND Collaboration.}
\end{flushright}

\clearpage
\renewcommand{\thepage}{\roman{page}}
\setcounter{page}{1}
\tableofcontents
\clearpage




\section*{Executive Summary}

An ancient relic of the Big Bang, neutrinos by the millions fill every cubic meter of space, a ghostly, unseen matrix in which the universe has evolved. Now, new experiments on these elusive particles are changing our understanding of the physical world. 

The first hint of the true nature of neutrinos was Nobel Prize winner Ray Davis's surprising discovery that fewer neutrinos come from the sun than were expected from our understanding of how the sun produces its energy. We now know that this is due to ``neutrino oscillations,'' a macroscopic consequence of the laws of quantum mechanics that govern the sub-atomic realm. Oscillations, in turn, tell us that neutrinos have mass, finally confirming a long-held suspicion. Since Davis's discovery, we have verified the existence of neutrino oscillations and neutrino mass using neutrinos produced in our atmosphere, in nuclear reactors, and by accelerators. 

We see the future of neutrino physics  framed in three overarching themes:

\begin{itemize}
\item {\bf Neutrinos and the New Paradigm:} Neutrinos have provided us with the first tangible evidence of phenomena beyond the reach of our theory of the laws of particle physics, the remarkably predictive ``Standard Model''. In the Standard Model, neutrinos do not have mass and do not oscillate. Through this crack in the edifice we are now peering, with no small excitement, to see the physics that lies beyond. It appears to be a glimpse of what physics is like at energies not seen since the Big Bang. Questions crowd upon us. 
The neutrino masses are not zero, but their values are uncertain by a factor of 100---what, exactly, are the masses? How much do neutrinos mix with each other, allowing one ``flavor'' of neutrino to change into another? Neutrinos, alone among matter particles, could be their own antiparticles. Are they? Our understanding of nature has been enormously enriched by the study of symmetry. Perhaps the most baffling symmetry is the `CP' symmetry (change particle to antiparticle and interchange left and right; everything should behave the same as before). Nature seems to have a bias here. Do neutrinos respect CP perfectly, a little, or not at all? We recommend the experimental program needed to build the foundations of the new paradigm.

\item {\bf Neutrinos and the Unexpected:}  Neutrino physics has been marked by ``anomalous,'' unexpected results that have proven to be absolutely correct and to have deep significance. Neutrinos may have even more extraordinary properties than those already seen. We have evidence for exactly three  flavors of neutrinos with normal interactions. Are there other flavors that lack these interactions? We describe an experimental program designed to be open to surprises. 

\item {\bf Neutrinos and the Cosmos:} Neutrinos originating from the Big Bang and from the cores of stars prompt us to find the connections between these particles and the universe. Neutrinos allow us to probe the origin and future of solar energy, upon which all life on earth depends. Understanding neutrinos is necessary to comprehend supernova explosions, perhaps the origin of the heaviest elements on earth. Neutrinos may have influenced the large-scale structure of the universe. Nature's bias with respect to CP is essential to explain why the universe contains matter but almost no antimatter. However, the bias seen in laboratory experiments outside the neutrino realm cannot solve this mystery. Perhaps neutrinos violate CP in a way that does help us solve it. We describe an experimental program to map out the connections between the neutrino and the cosmos. 
\end{itemize}

While the questions to be answered are clear, the best strategy  demands thoughtful planning. Developing the strategy is made more challenging by the fact that the field spans the studies of particle physics, nuclear physics, astrophysics, and particle beams. Drawing on the wide-ranging expertise of members of the neutrino community in these areas, we report the results of our study on the future of neutrino physics, organized by four Divisions of the American Physical Society. A central purpose of this report is to communicate to U.S. decision-makers the consensus that has emerged among our group on three recommendations: 
\begin{flushleft}
\frame{
\hspace{0.15 in}\begin{minipage}[]{6.2 in}
\vspace{0.1 in}
\noindent
\textcolor{blue}{{\bf We recommend, as a high priority, a phased program of sensitive searches for neutrinoless nuclear double beta decay.}} {\it In this rare process,  one atomic nucleus turns into another by emitting two electrons.  Searching for it is very challenging, but the question of whether the neutrino is its own antiparticle can only be addressed via this technique. The answer to this question is of central importance, not only to our understanding of neutrinos, but also to our understanding of the origin of mass.} 

\bigskip
\textcolor{blue}{{\bf We recommend, as a high priority, a comprehensive U.S. program to complete our understanding of neutrino mixing, to determine the character of the neutrino mass spectrum and to search for CP violation among neutrinos.}} {\it This comprehensive program would have several components: an experiment built a few kilometers from a nuclear reactor, a beam of  accelerator-generated neutrinos aimed towards a detector hundreds of kilometers away, and, in the future, a neutrino `superbeam' program utilizing a megawatt-class proton accelerator. The interplay of the components makes possible a decisive separation of neutrino physics features that would otherwise be commingled and ambiguous. This program is also valuable for the tools it will provide to the larger community. For example, the proton accelerator makes possible a wide range of research beyond neutrino physics.} 

\medskip

{\it The development of new technologies will be essential for further advances in neutrino physics. On the horizon is the promise of a neutrino factory, which will produce extraordinarily pure, well-defined neutrino beams. Similarly challenging are the ideas for massive new detectors that will yield the largest and most precise samples of neutrino data ever recorded. These multipurpose detectors can also be used for fundamental and vitally important studies beyond the field of neutrino physics, such as the search for proton decay.  } 

\bigskip
\textcolor{blue}{{\bf  We recommend development of an experiment to make precise measurements of the low-energy neutrinos from the sun.}} {\it So far, only the solar neutrinos with relatively high energy, a small fraction of the total, have been studied in detail. A precise measurement of the low-energy neutrino spectrum  would test our understanding of how solar neutrinos change flavor, probe the fundamental question of whether the sun shines only through nuclear fusion, and allow us to predict how bright the sun will be tens of thousands of years from now.}
\vspace{0.1 in}
\end{minipage}
\hspace{0.1 in}
}
\end{flushleft}
These recommendations are made in the context of certain assumptions about the groundwork for the new experimental program. The assumptions include:

\begin{itemize}
\item {\bf Continuation and strong support of the existing program.} The future program we recommend depends on successful completion of the investigations now in progress.  We have identified four areas to address: continued increase in the proton intensity
for neutrino experiments at Fermilab, resolution of an experimental indication of neutrino flavor change over short distances,  measurement of solar
neutrinos of intermediate energy, and continued support of R\&D for detection of ultra-high energy
astrophysical neutrinos. With these and other modest improvements, the current phase of the neutrino program can be accomplished.

\item {\bf Underground laboratory facilities.}  The extreme rarity of neutrino interactions  requires that  experiments that are central to our proposed program, including double beta decay, studies with the multipurpose very large detector, and solar neutrino research, be carried out deep underground in appropriately designed laboratories.

\item {\bf Determination of the neutrino reaction and production cross sections required for a precise understanding of neutrino-oscillation physics and the neutrino astronomy of astrophysical and cosmological sources.} Our broad and exacting program of neutrino physics is built upon precise knowledge of how neutrinos interact with matter.

\item {\bf Research and development to assure the practical and timely realization of accelerator and detector technologies critical to the recommended program.} Of particular importance are R\&D efforts aimed toward development of a high-intensity proton driver, a neutrino factory, a very large neutrino detector, and techniques for detection of ultra-high-energy neutrinos.

\item {\bf International cooperation.} We advocate that the program to answer the outstanding neutrino questions be international. In this report, we recommend a U.S. program that will make unique contributions to this international effort, contributions that will not be duplicated elsewhere. The U.S. program, involving experiments within the U.S. and American participation in key experiments in other countries, has the potential to become the best in the world. But it must cooperate with the programs of other nations and regions. The programs to be carried out throughout the world must complement each other. We explain how they can do this.

\end{itemize}

The experimental program described in this study is intended to be a very fruitful investment in fundamental physics. The selection is physics-rich, diverse, and cost-effective. A timeline has been developed to synchronize aspects of the program and to be integrated with the worldwide effort to reach an understanding of the neutrino. The program components are chosen to provide unique information and thereby enhance companion studies in high energy physics, nuclear physics, and astrophysics. 
There are rare moments in science when a clear road to discovery lies ahead and there is broad consensus about the steps to take along that path.   This is
one such moment.

\renewcommand{\theenumi}{\arabic{enumi}}

\clearpage
\end{onecolumn}
\begin{twocolumn}
\renewcommand{\thepage}{\arabic{page}}
\setcounter{page}{1}

\section{Introduction}

We live within a matrix of neutrinos.   Their number far exceeds
the count of all the atoms in the entire universe.  Although they
hardly interact  at all, they helped forge the elements
in the early universe, they tell us how the sun shines, they may even cause the titanic explosion of a dying star.  They may well be the reason we live in a universe
filled with matter -- in other words, a reason for our being here.

Much of what we know about neutrinos we have learned in just the last six years.  Neutrino discoveries
have come so fast we have barely had time to rebuild the conceptual
matrix by which we hope to understand them.

The new discoveries have taught us two important things: that
neutrinos can change from one type to another; and that, like other
fundamental particles of matter, they have mass.  The implications of these new facts reach well
beyond just neutrinos, and affect our understanding of the sun, our
theory of the evolution of the Universe, and our hope of finding a
more fundamental theory of the subatomic world. We now have so many
new questions, our task in this Study has been especially difficult.
We are most certain of one thing:  neutrinos will continue to
surprise us.

\vspace{0.2in}

\noindent {\large {\it The Story:}}

\vspace{0.1in}

A crisis loomed at the end of the 1920's -- a decade already filled
with revolutions.  One of physics' most sacred principles -- the
conservation of energy -- appeared not to hold within the subatomic
world.  For certain radioactive nuclei, energy just seemed to
disappear, leaving no trace of its existence.

In 1930, Wolfgang Pauli suggested  a ``desperate way out.''  Pauli postulated that the missing energy was
being carried away by a new particle, whose properties were such that
it would not yet have been seen: it carried no electric charge and scarcely
interacted with matter  at all.  

Enrico Fermi soon was able to show that while the new particles
would be hard to observe, seeing them would not be impossible. What was
needed was an enormous number of them, and a very large detector.
Fermi named Pauli's particle the neutrino, which means `little neutral
one'.  More than two decades after Pauli's letter proposing the neutrino, Clyde Cowan and
Fred Reines finally observed (anti)neutrinos emitted by a nuclear
reactor.  Further studies over the course of the next 35 years taught
us that there were three kinds, or `flavors,'  of neutrinos
(electron neutrinos, muon neutrinos, and tau neutrinos) and that, as
far as we could tell, they had no mass at all.  The neutrino story
(Fig.~\ref{fig:excitement}) might have ended there, but developments
 in solar physics changed everything.

\begin{figure}[h]
\begin{center}
\includegraphics[height=2in]{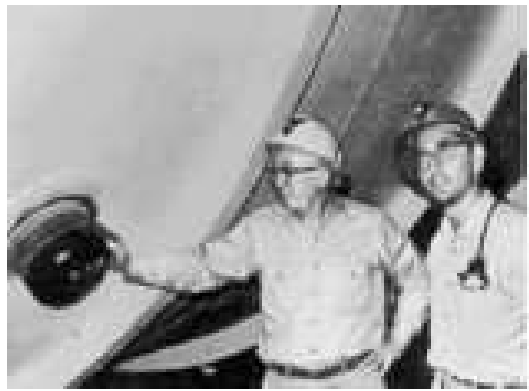}
\caption{Ray Davis (left) and John Bahcall with the first solar neutrino 
detector in the Homestake mine.}.
\label{fig:davis}
\end{center}
\end{figure}
\begin{figure*}[t]
\begin{center}

\includegraphics[height=5in]{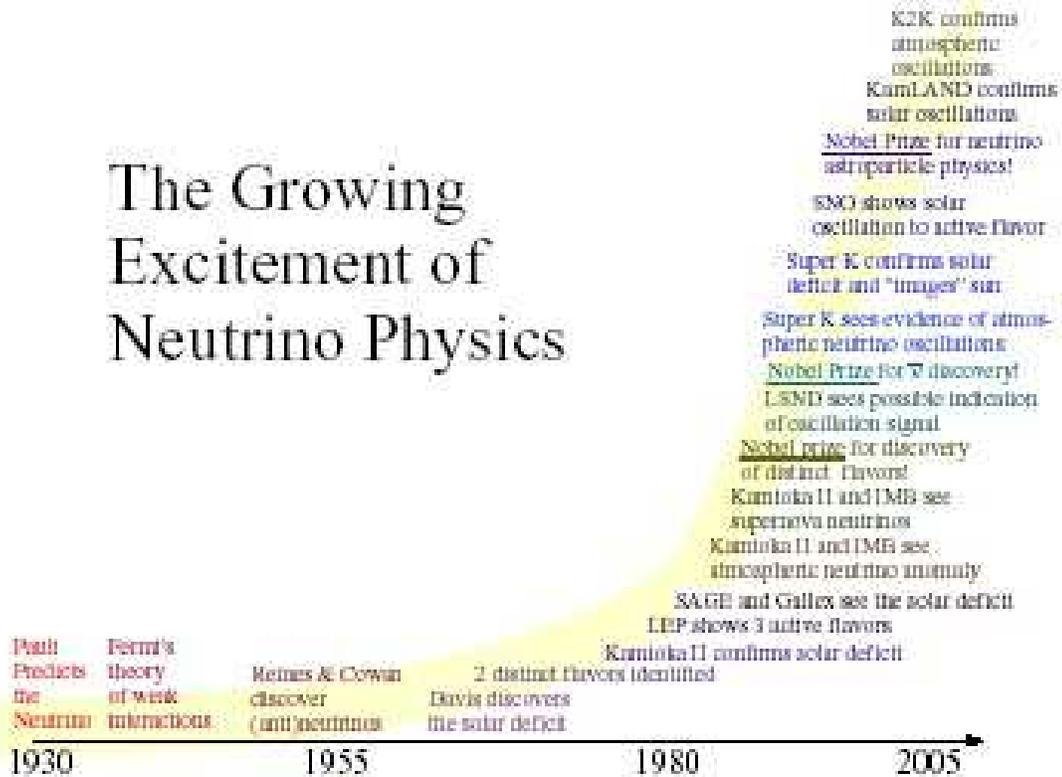}

\caption{Important events that have led to the present excitement in 
neutrino physics.}.

\label{fig:excitement}

\end{center}
\end{figure*}

In 1919, Sir Arthur Eddington had suggested that the sun's
multi-billion year age could be explained if its power source was the
``well-nigh inexhaustible'' energy stored in atomic nuclei.  With Fermi's neutrino theory, Hans
Bethe and Charles Critchfield in 1938 created the first detailed theory
of the nuclear furnace burning in the sun's core.

Neutrinos are produced in great numbers by those nuclear reactions, and 
can pass from the solar center to us directly.  While
the light we see from the sun represents energy created in the core tens
of thousands of years ago, a neutrino created in the sun right now
will reach us in just over eight minutes.  But if neutrinos can pass
easily through the sun, how could we possibly detect them on Earth?
In the mid-1960's experimentalist Raymond Davis, Jr.~and theorist John Bahcall
thought about this problem.  Bahcall's detailed calculations showed
that there might just be enough neutrinos produced in the sun that
they could be observed on earth, and Davis set out to build a detector
that could see the neutrinos.  His detector weighed hundreds of tons, and he had
to be able to detect the few atoms each week that had been transformed
by neutrinos.  What Davis saw was surprising.

While he did observe neutrinos, Davis found  only roughly 1/3 the
number Bahcall had predicted.  Davis' experiment was exceedingly
difficult, and Bahcall's calculations equally so.  Many physicists
believed that it was likely that either, or perhaps both, were in
error.  But over the next three decades, solar neutrino predictions
became more refined, and new experiments invariably saw fewer than
predicted.  The mystery would not go away.

 \begin{figure}[p]
\framebox{
\hspace{0.1 in}\begin{minipage}[]{6.5 in}
\vspace{0.1 in}
\noindent
{\bf \LARGE Neutrinos in a Nutshell} \newline

Neutrinos are the most abundant matter particles, called ``fermions,'' in the universe. 
Unlike their relatives, the electron and the quarks, they have no 
electrical  charge.  

\medskip

There are three different types (or `flavors') of electron-like particles,
each with a different mass:  the electron ($e$) itself, the muon ($\mu$)
weighing 200 times more than the  electron,  and the tau ($\tau$) which
weighs 18 times more than the muon.   For each of these charged particles there is also a neutrino. Collectively, these six particles ($e$, $\mu$, $\tau$, $\nu_1$, $\nu_2$ and $\nu_3$) are known as 
the `leptons',  which comes from the Greek word meaning `thin',  
`subtle', or `weak'.

\includegraphics[width=3in]{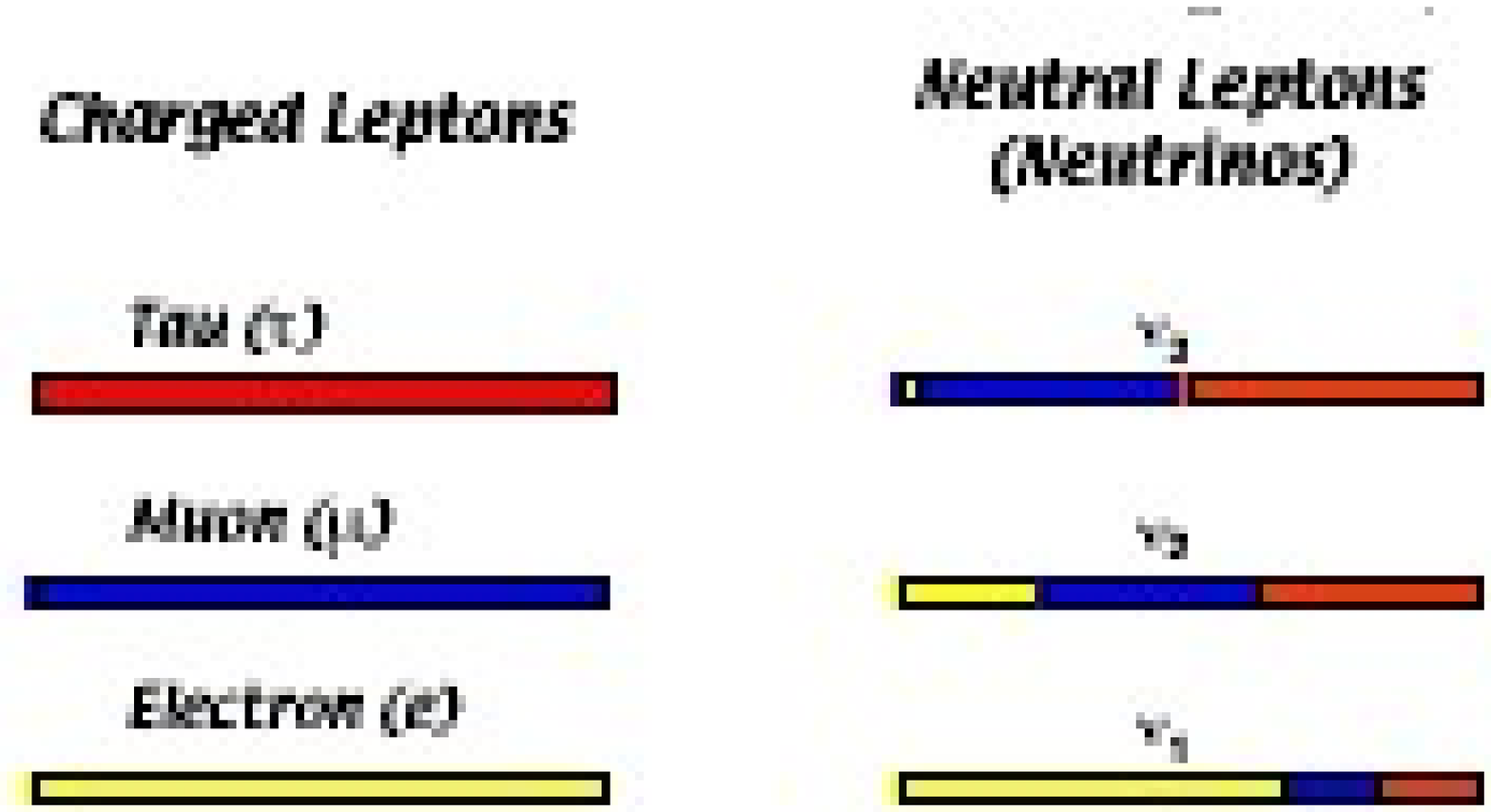}
\hspace{0.2 in}
\begin{minipage}[b]{3 in} 
The leptons.  The colors indicate the `flavors' of 
the charged leptons, electron, mu, and tau. The flavors determine 
what happens when a lepton collides with another particle.
\bigskip
\end{minipage}
\label{fig:leptons}

Neutrino masses are exceedingly tiny, compared to the masses of their 
charged brethren. It is only from discoveries made in the last six 
years that we know that these masses are not exactly zero, and that the heaviest 
of them must have a mass at least one ten-millionth the electron's mass. 
Moreover, we know that the masses are all different.

\medskip
Like all the other particles of matter, neutrinos have antimatter partners,
denoted with a bar on top: {\it e.g.} $\bar \nu_1$, $\bar \nu_2$, $\bar \nu_3$.
Unlike any other fermion, though,  the 
$\nu$ and $\bar \nu$ may in fact be the same particle.  

\medskip
Drawn six years ago, the figure above would have the neutrinos each with a single, different flavor, like the charged leptons.  Neutrinos are created with other particles through a force 
appropriately named the `weak interaction,' and the weak interaction 
does not change flavor.   For example, in the beta decay studied by Pauli in 1930 the weak interaction makes an antielectron and an `electron neutrino,'  $\nu_e$.  A weak interaction that made an antimuon would also make a `mu neutrino,' $\nu_\mu$, and so forth.  But what are those `particles'?  The only way nature can construct a neutrino that is totally electron flavored is to form a quantum-mechanical mixture of exactly the right amounts of the mixed-flavor particles $\nu_1$, $\nu_2$ and $\nu_3$.  What had always been thought of as a simple particle,  $\nu_e$ is actually a quantum-mechanical Neapolitan of the 3 neutrinos with definite masses.

%
\medskip
As time passes, or the neutrino
travels, the quantum waves that accompany the different parts get out of step because the masses are different. Depending on the distance travelled, what was originally produced
as an electron flavored `neutrino' can become mu flavored or
tau flavored as the components shift. This is the phenomenon called neutrino
oscillations, and it provides our best evidence that neutrinos have
distinct, nonzero masses.

\medskip
There is a lot still to learn about the masses and flavors.  We are
now trying to measure the flavor contents of each neutrino, and we
represent them by 3 trigonometric angles called $\tha$, $\thb$, and
$\thc$.    The masses themselves are only known within broad
ranges, although oscillations tell us quite a lot about the
differences.

\vspace{0.1 in}
\end{minipage}
\hspace{0.1 in}
}
\end{figure}

\clearpage

The best explanation that encompassed both the theoretical prediction
and the experimental results was that the neutrinos produced in the
sun were changing from one flavor to another.  Experiments like Davis'
were sensitive only to electron neutrinos, the only kind the sun can
produce.  If, on their way from the sun to the earth, some of the
electron neutrinos changed into the other flavors, they could sail
through the detectors completely unobserved.  Neutrinos of the sort envisaged by Pauli and enshrined in our Standard Model of particles could not perform this feat.

While physicists puzzled over the solar neutrino experiments, a new
neutrino mystery arose in the mid 1980s.  When cosmic rays hit the
earth's atmosphere, they create showers of other particles, including
neutrinos.  The Kamiokande and IMB experiments,  built to search for proton decay,  found that the number of
muon neutrinos created in the atmosphere appeared to be smaller than
expected.  The experimenters pointed out that, like the solar
neutrinos, this could be true if the muon neutrinos were actually
changing into undetected neutrinos, in this case tau neutrinos.  But the
experiments were very difficult, and many physicists again attributed
the deficit to error.

Our consistent picture has been the result of careful testing and repetition of important experiments.  A recent experimental indication that neutrinos and antineutrinos are the same particle, as is anticipated on theoretical grounds, will require confirmation. One experimental observation does not fit neatly into the picture of 3 active neutrinos that mix and have mass. In the LSND experiment, muon antineutrinos appear to convert to electron antineutrinos over a short path.  The observation is being checked in a new experiment, MiniBOONE.

The discovery of neutrino flavor transformation and mass answered questions that had endured for dec\-ades.  As those veils have lifted, burning new questions about the physical and mathematical neutrino matrix challenge us.

\begin{centering}
\begin{flushright}
\vspace{3 in}
{\em Facing Page: Inside the SNO Detector. \\
The technician is crouching on the floor \\
of a 12-m diameter acrylic sphere so \\
transparent it can hardly be seen. \\
Some 10,000 photomultipliers \\ 
surround the vessel. \\
\tiny Photo: Lawrence Berkeley National Laboratory}
\end{flushright}
\end{centering}
\clearpage
\thispagestyle{empty}
\includegraphics[height=8.9in]{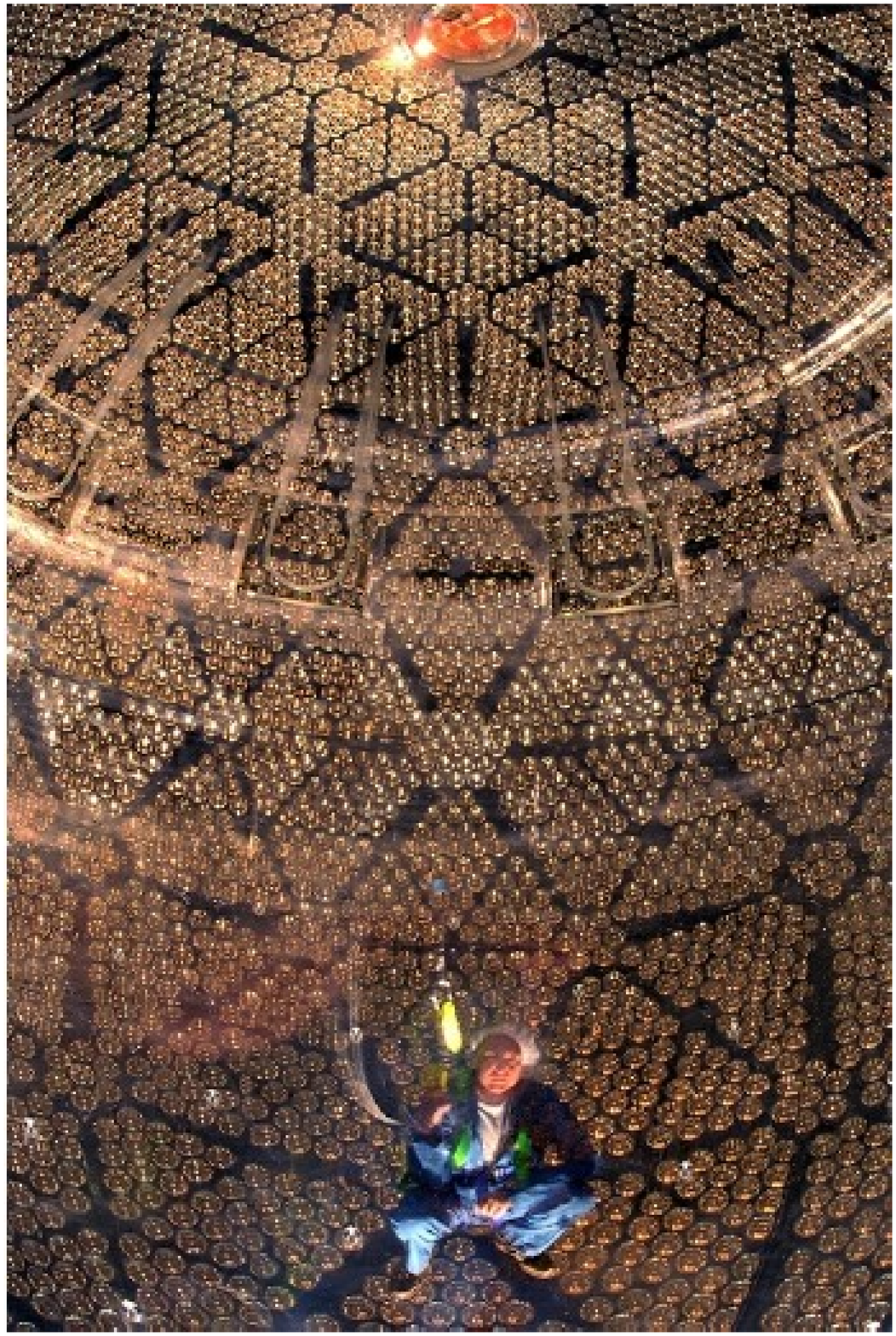} \\
\clearpage

\section{Answers and Questions}

The story of neutrinos continues to be written.  As the narrative
unfolds, three themes have crystallized that broadly define the
science.  Within each of these themes, we are confronted by basic
questions.  Understanding the nature of neutrinos has become a
critical issue at the frontiers of physics, astrophysics, and
cosmology.  There is universal agreement about the questions that must
be answered.  It is only the difficulty of obtaining the answers that
requires a well planned strategy.

\subsection{Neutrinos and the New Paradigm}

The neutrino discoveries of the last decade force revisions to the basic picture of
the elementary particles and pose a set of well-defined but presently
unanswered questions,  questions of fundamental importance.

\vspace{0.2 in}\vbox{
{\bf \noindent $\bullet$ \textcolor{red}{What are the masses of the 
neutrinos?}} 

\begin{flushleft}
\frame{\hspace{0.1 in}
\begin{minipage}[]{2.8 in}
\begin{centering}
\vspace{0.1 in}
{\bf The Mass that Roared}

{\it The  discovery that neutrinos have mass is a breakthrough.  For 30 years the trustworthy, ``Standard'' model has unfailingly been able to describe anything in the particle world, in some cases to 10 decimal places. That model asserts that neutrinos {\it are massless}.  Physicists expected that one day the model would fail, even hoped for it, because the model appears to be a simplification of a more complete description of nature.
  }\vspace{0.1 in}
\end{centering}
\end{minipage}\hspace{0.1 in}
}\end{flushleft}}
  
  The combination of solar, atmospheric, accelerator, and reactor
  neutrino data reveals that the flavor change is due to a quantum phenomenon called ``oscillations''  and shows that at least two neutrinos
  have nonzero, distinct masses. This simple fact has forced us to modify our description of particle physics, the ``Standard Model,'' for the first time since it was created over 25 years ago.    If there are three neutrinos with
masses $m_1$, $m_2$, and $m_3$, oscillation experiments
give the differences between the squares of the masses.
We express these as $\Delta m_{12}^2$, which is $m_2^2-m_1^2$,
$\Delta m_{23}^2$, which is $m_3^2-m_2^2$, and $\Delta m_{13}^2$ which is 
$m_3^2-m_1^2$.   One can see that any two difference pairs, sign and magnitude, are sufficient to fix the third.
  
Oscillations tell us about mass differences, but what about the masses themselves?  In the laboratory, 
precise measurements of the tritium beta-decay spectrum constrain
the average of the three neutrino masses to be less than 2.2~eV.  For comparison
(Fig.~\ref{fig:allmasses}), the electron, the lightest of the charged
elementary particles, has a mass of 510,999 eV.  But the
oscillation results point to an average neutrino mass not smaller than 0.02~eV.
The mass is boxed in: it must lie between 0.02 and 2.2 eV.  

Interestingly, studies of the large-scale structure of the visible
universe combined with the precise determination of the cosmic
microwave background radiation from experiment put the
average neutrino mass at less than 0.5~eV. Now we must pin it
down.
\begin{figure}
\hspace{-0.3in}
\includegraphics[width=3.3in]{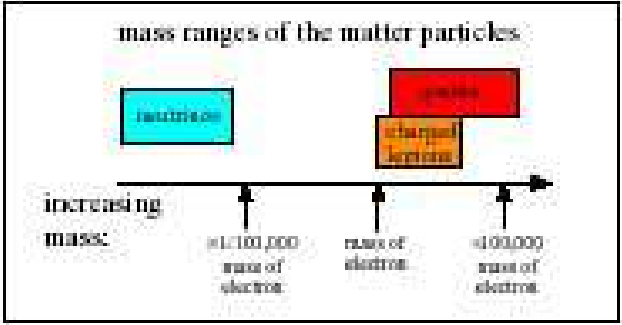}
\caption{The masses of the fundamental fermions, the particles of matter. The masses are shown on 
  a log scale.  The individual neutrino masses are not well known, and the average is only known to within two
  orders of magnitude.  There is a  surprisingly large difference between
  the neutrino masses and the masses of the quarks and charged
  leptons.}
\label{fig:allmasses}
\end{figure}
There are three kinds of experiments focused on establishing the absolute value of the neutrino masses:
\begin{enumerate}

\item precise  experiments on the beta decay of tritium,
  seeking to directly measure the average neutrino mass.

\item neutrinoless double-beta-decay experiments, which have
  sensitivity to another linear combination of neutrino masses,
  provided that neutrinos are their own antiparticles; and

\item precision studies of the distribution of the cosmic microwave
  background combined with observations of the large-scale structure of the
  universe revealed by clusters of galaxies.
\end{enumerate}

The two possible orderings of the masses, or hierarchies, are depicted
in Fig.~\ref{fig:masses1}, and are often referred to as  ``normal'' and  ``inverted.''  We currently do not know which is correct.  Knowing the
 ordering of the neutrino masses is 
important. For example, in the case of an inverted hierarchy,
there are at least two neutrinos that have almost the same mass to the
one percent level.  We have yet to encounter two different
fundamental particles with nearly identical masses.  If neutrinos have this property, it surely points to a new and fundamental aspect of  Nature.

\begin{figure*}
\begin{center}
\includegraphics[width=5in]{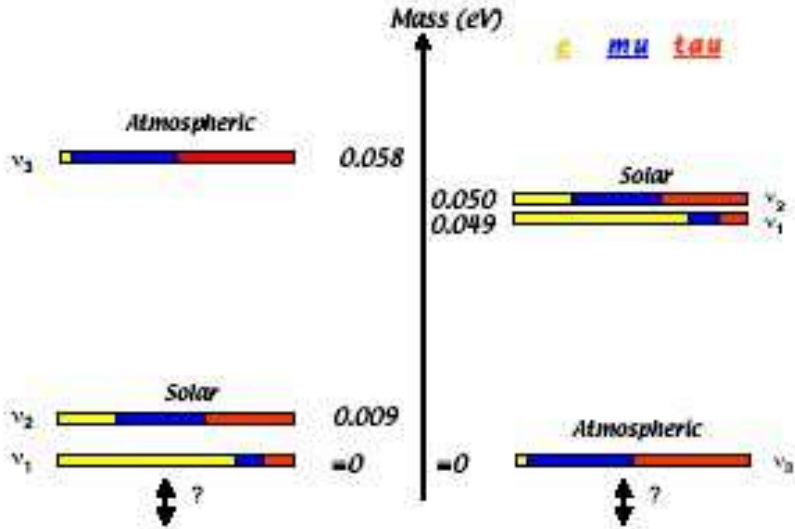}
\caption{The two
  possible arrangements (hierarchies) of the masses of the three known
  neutrinos, based on oscillation data.  The picture shows the situation for the mass of the lightest
  neutrino being zero; but in fact, from experiment, the
  average mass of the neutrinos may be as large as 2.2 eV. On the left is a ``normal'' hierarchy, and on the right an ``inverted'' one.}
\label{fig:masses1}
\end{center}
\end{figure*}

Future neutrino experiments may determine the neutrino
mass hierarchy. Two techniques have the potential of determining the hierarchy:
\begin{enumerate}
  
\item accelerator-based long-baseline oscillation experiments with
  baselines in the vicinity of 1000 km or more; and

\item very large atmospheric neutrino experiments that can
  independently measure the oscillation of neutrinos and
  antineutrinos.
\end{enumerate}

\vspace{0.2 in}\vbox{
{\bf \noindent $\bullet$ \textcolor{red}{What is the pattern
of mixing among the different types of neutrinos?}}

\begin{flushleft}
\frame{\hspace{0.1 in}
\begin{minipage}[]{2.8 in}
\begin{centering}
\vspace{0.1 in}
{\bf Double Identity}

{\it Neutrinos exist with a dual identity.  The neutrinos with definite mass are not the
  objects we thought we knew, $\nu_e$, $\nu_\mu$, and $\nu_\tau$.
  They are particles, $\nu_1$, $\nu_2$, and $\nu_3$, each with a rainbow
  of the three flavors.  The connection between the dual
  identities, which is manifested in the phenomenon of neutrino oscillations,  is a  key
  to the physics beyond the Standard Model.  }\vspace{0.1 in}
\end{centering}
\end{minipage}\hspace{0.1 in}
}

\end{flushleft}
}

 Mathematically, we
  relate the neutrinos with definite mass to the flavors via a mixing matrix.
  The same phenomenon is observed for the quarks, and several
  decades of research have gone into measuring and interpreting what is referred to as the ``CKM quark mixing matrix.''

Like the neutrinos, the quarks have mass states that have
mixtures of flavor.    One would think that we could look to the quarks to
understand the neutrinos, but the theoretical analogy proves unhelpful.    Unlike the numbers that describe quark
mixings, which are small, the mixing of the neutrinos
is large.  The origin of this striking difference is not presently understood.

We can describe the mixing in ``matrix notation:'' 

\noindent {\bf Neutrinos}
$$U_{MNSP}\sim
\left(
\begin{array}{ccc}
 \mbox{\Huge 0.8} &   \mbox{\Huge 0.5}  &   \mbox{\large \bf ?} \\
  \mbox{\Huge 0.4} &   \mbox{\Huge 0.6}  &  \mbox{\Huge 0.7} \\
  \mbox{\Huge 0.4} &   \mbox{\Huge 0.6}  & \mbox{\Huge 0.7}
\end{array}
\right)
$$
{\bf Quarks}
$$V_{CKM}\sim
\left(
\begin{array}{ccc}
 \mbox{\Huge 1} &   \mbox{\large 0.2}  &   \mbox{\tiny $0.005$} \\
  \mbox{\large 0.2} &   \mbox{\Huge 1}  &  \mbox{\small $0.04$} \\
  \mbox{\tiny $0.005$} &   \mbox{\small $0.04$}  & \mbox{\Huge 1}
\end{array}
\right)$$
 The difference between the large numbers
dominating the neutrino matrix and the small numbers for the quark matrix
is dramatic.

Determining  all the elements of the neutrino mixing matrix is
important because it is likely that, in a way we do not yet understand, they contain fundamental
information about the structure of matter.   We see mixing in other contexts in physics, and it generally is a result of the interaction of simpler, more primitive, systems.  The mu and tau flavors,
for example, may in fact  be mixed as much as is possible -- is it so,
and, if so, why?

\begin{figure*}[t]
\begin{center}
\includegraphics[width=6.5in]{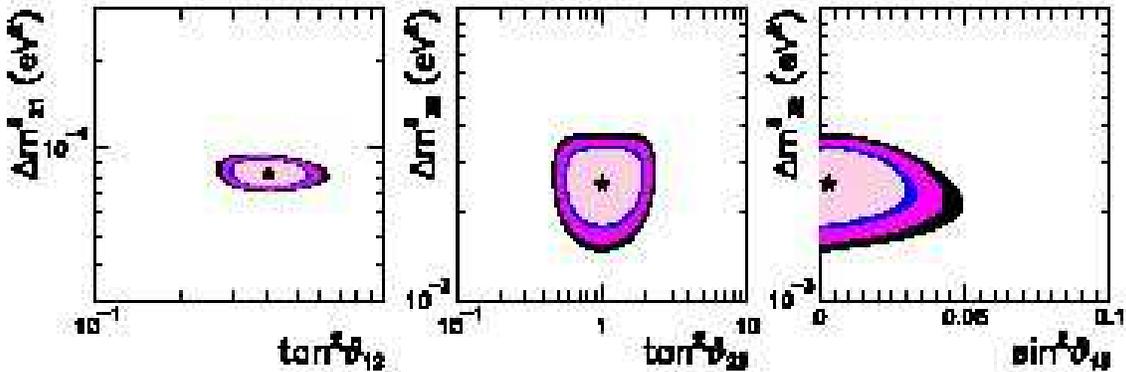}
\caption{Current experimental constraints on the three mixing angles, $\tha$, 
$\thb$, and $\thc$, and their dependence on the two known 
mass-squared differences, $\dma$ and $\dmc$.  The star indicates the most
likely solution.  The contours correspond 
to certain confidence levels that the parameter pairs lie within. }
\label{fig:islands}
\end{center}
\end{figure*}

For three neutrino species, the neutrino mixing matrix {\bf U} has
nine elements, but all of them are determined by the same four 
or six underlying quantities -- six if neutrinos are their own 
antiparticles, four otherwise. These underlying quantities are three 
mixing angles:  the ``solar angle''
$\theta_{12}$, the ``atmospheric angle'' $\theta_{23}$, and
$\theta_{13}$; and one or three complex phases.  Neutrino mixing and
mass together lead to neutrino oscillations --- this is how we learned
that neutrinos have mass --- and the detailed study of the oscillation
phenomenon allows us to measure the three mixing angles and one of
the CP-violating phases, referred to as $\delta$.

We can describe the mass states and neutrino mixings using the set of
bars in Fig.~\ref{fig:masses1}.  Each bar represents a
neutrino of a given mass, $\nu_1$, $\nu_2$, and $\nu_3$.  We use mixing angles to describe how much of each flavor (electron, muon, or tau) can be found in each 
neutrino.  In this diagram we denote the fractional flavors by the
color in the bar.  Yellow is electron flavor, blue is muon flavor, and
red is tau flavor.  For concreteness we have picked certain flavor
fractions for each bar, although the  fractional amounts
are presently  known imprecisely or not at all.

We can now connect the diagram of Fig.~\ref{fig:masses1} to the mixing
angles we measure:
\begin{itemize}
\item   $\sin^2\theta_{13}$ is equal to the amount of
$\nu_e$ contained in the $\nu_3$ state (the yellow in the $\nu_3$
bar).  
\item $\tan^2\theta_{12}$ is equal to the amount of $\nu_e$ in
$\nu_2$ divided by the amount of $\nu_e$ in $\nu_1$, {\it i.e.}, the
ratio of the yellow fraction of the $\nu_2$ bar to the yellow
fraction of the $\nu_1$ bar in Fig.~\ref{fig:masses1}. We currently
know that $\tan^2\theta_{12}<1$, which means that there is more
$\nu_e$ in $\nu_1$ than in $\nu_2$. 
\item $\tan^2\theta_{23}$ is
the ratio of $\nu_{\mu}$ to $\nu_{\tau}$ content in $\nu_3$, {\it
  i.e.,} the fraction of the $\nu_3$ bar in Fig.~\ref{fig:masses1}
colored blue divided by the fraction colored red. We currently do not
know whether the $\nu_3$ state contains more $\nu_{\mu}$ or more
$\nu_{\tau}$, or an equal mixture.
\end{itemize}
Figure \ref{fig:islands} summarizes our experimental knowledge of the 3 mixing angles.  The differences of the squared masses provide enough information now at least to link  together the masses of the 3 known neutrinos for the first time. Two of the angles are large.  The ``solar angle'' is now fairly well determined from experiment: $\tha = 32.3\pm1.6^o$. The ``atmospheric angle'' is not as accurately known, but appears to be as large as it can be:  $\thc = 45\pm 8^o$.  The third angle, $\thb$, is known only to be relatively small, less than $10^o$.  That is a major obstacle.  Not only do we not yet have a complete picture of the pattern of mixing, but if this angle is zero, there is then no possibility of testing whether the important ``CP symmetry'' is preserved or violated by neutrinos (see below).   What new experiments can improve our knowledge of the 3 angles, especially $\thb$? 
\begin{enumerate}

\item Precision solar neutrino experiments;

\item Very precise measurements, at the 1\% level or better, of the
  flux and spectrum of electron-flavor antineutrinos produced in
  nuclear reactors and observed a few kilometers away from the
  source;
  
\item Accelerator-based long-baseline oscillation experiments with
  baselines of hundreds of km or more.
\end{enumerate}
 
\vfill\eject
 {\vspace{.2 in}\vbox
{\bf \noindent $\bullet$ \textcolor{red}{Are 
neutrinos their own antiparticles?}} 

\begin{flushleft}
\frame{\hspace{0.1 in}
\begin{minipage}[]{2.8 in}
\begin{centering}
\vspace{0.1 in}
{\bf Neutrino and Antineutrino}

{\it Particles and antiparticles  have opposite charges.  What are we to
  make of neutrinos, which have no charge?  Is there anything that
  requires a distinction between neutrinos and antineutrinos?  If
  there is not, perhaps neutrinos and antineutrinos are really the
  same particle.  This possibility can explain why neutrinos are so
  light, yet not completely massless, and it also points to the
  existence of neutrinos so heavy we cannot possibly make them in the
  laboratory.  Intriguingly, their mass-energy happens to be about the same as the
  energy where the known forces (except gravity) may unite as one. We
  must find out if neutrinos are their own antiparticles.
}\vspace{0.1 in}
\end{centering}
\end{minipage}\hspace{0.1 in}
}
\end{flushleft}
}
  The requirement that Albert Einstein's theory of special relativity
  also be applicable to the weird world of quantum mechanics led to
  the remarkable prediction, by Paul A. M. Dirac, that for every
  particle there exists an antiparticle. The particle and the
  antiparticle have identical mass and spin. 
  However, they have opposite electric 
charges, and any other charge-like attributes they may possess. Neutral 
particles are special in that they can be their own antiparticles. This is 
true of several neutral particles that are not fermions, including the 
photon -- the particle of light.

Neutrinos are the only elementary neutral fermions known to exist.  Being neutral, they 
could also be their own antiparticles. Now that we know neutrinos have
mass, we can address this most fundamental question.  The answer to this question is needed in order to build a New Standard Model. There are two
completely different ways of ``adding'' massive neutrinos to the old
Standard Model -- one that allows neutrinos to be their own
antiparticles, and one that does not -- and we must know which one is
correct in order to proceed. As things stand, we no longer can claim we know the
equations that describe all experimentally observed phenomena in
particle physics.

In practice, we attack this problem by asking what must be true if the
neutrinos are {\sl not} their own antiparticles. If the neutrino and
antineutrino are distinct particles, they must possess some new
fundamental ``charge'' which distinguishes the neutrino from the
antineutrino. This charge is called ``lepton number.'' We assign the
neutrinos and the negatively charged leptons lepton number $+1$,
and the antineutrinos and the positively charged leptons
lepton number $-1$. If lepton number is violated by {\it any} physical process,
it would not be a conserved charge. This necessarily would imply that
the neutrinos {\it  are} their own antiparticles. If, on the other hand,
lepton number is always conserved, it reveals a new fundamental
symmetry of Nature, one we did not know existed before.
\footnote{Note to experts: lepton number is known to be violated in
  the Standard Model by nonperturbative effects. One should replace
  everywhere `lepton number' by `baryon number minus lepton number'
  ($B-L$), which is the non-anomalous global symmetry of the old
  Standard Model Lagrangian. If it turns out that neutrinos are not
  their own antiparticles, we are required to ``upgrade'' $B-L$ from
  an accidental symmetry to a fundamental one.}

Currently, we have no confirmed experimental evidence that lepton
number is violated.  By far the most sensitive probe of lepton number
violation is neutrinoless double beta decay.  In that process, related to the beta decay process discussed earlier in which a single
neutron decays to a proton, an electron, and an antineutrino,
 it may be energetically favorable for two
neutrons to beta decay simultaneously.  This process, called double
beta decay, occurs rarely; it results in two antineutrinos, two electrons
and two protons.  If neutrinos are their own antiparticles, then, in
principle, the antineutrino pair could annihilate, resulting in
neutrinoless double beta decay: One 
nucleus decays into another nucleus plus two electrons, thereby violating 
lepton number by two units.

The outcomes of future searches for neutrinoless double beta decay,
combined with results from neutrino oscillation experiments and direct
searches for neutrino masses, may not only unambiguously determine
whether the neutrino is  its own antiparticle, but may also constrain the neutrino masses themselves.

{\bf \noindent $\bullet$ \textcolor{red}{Do neutrinos 
violate the symmetry CP?}} 

\begin{figure}[h!]
\begin{center}
\includegraphics[width=3in]{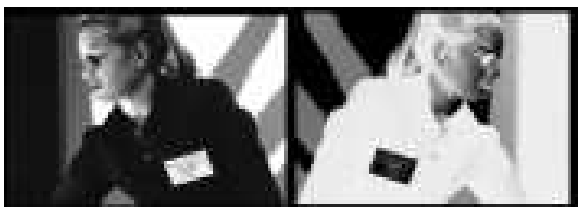}
\caption{A neutrino physicist seen in a CP Mirror, which
inverts spatially and maps matter to antimatter. CP invariance implies
the same behavior for both sides of the mirror.}
\label{fig:CPwoman}
\end{center}
\end{figure}
\vspace{-0.1in}
\vbox{
\begin{flushleft}
\frame{\hspace{0.1 in}
\begin{minipage}[]{2.8 in}
\begin{centering}
\vspace{0.1 in}
{\bf The Mirror Cracked}

{\it When you look at yourself in a mirror, you see a perfect spatial
  reflection that behaves just as you do, only in reverse.  Nature's
  particle mirror, which we call ``CP,'' is one that reflects not only
  in space, but from matter to antimatter.  This particle mirror is
  known to have a tiny flaw: at a very small level quarks don't behave
  like their looking-glass partners.  But what is small for quarks
  could be large for neutrinos, and through this crack in Nature's
  mirror, we may see physics far beyond the present energy scales.
}\vspace{0.1 in}
\end{centering}
\end{minipage}\hspace{0.1 in}
}
\end{flushleft}
}
CP invariance says that when matter is mirrored spatially and then
  converted to antimatter, the result should behave identically to the
  original particle (see Fig.~\ref{fig:CPwoman}).  Guided by the quark
  sector, though, we expect CP-invariance to be violated in the
  neutrino sector at a small level.  We are also led to conclude that,
  as in the quark sector, several CP-invariance violating phenomena in
  neutrino physics should be described in terms of the same
  fundamental parameter --- the CP-violating phase $\delta$ contained
  in the mixing matrix.  We have learned, however, that the guidance
  provided by the quark sector and other ``theoretical prejudices''
  can lead us astray. 
There is no fundamental reason to believe that
  the mechanism for neutrino CP-invariance violation is the same as
  the one observed in the quark sector.  Only experiments can determine the size of CP-invariance violation among neutrinos.

A prerequisite, as we have mentioned, for being able to observe CP-invariance violation in the neutrino sector is that the third mixing angle, $\thb$,  not be vanishingly small.  Experimentally one  must be simultaneously sensitive to the effects of all three mixing angles in order to see CP-violating phenomena. Given that fact, the best, and only practical, approach is  accelerator-based long-baseline
oscillation experiments. One test is to compare
electron-flavor to muon-flavor neutrino oscillations to
electron-flavor to muon-flavor antineutrino oscillations.  A difference would be a
CP-invariance violation, although in practice the presence of matter can counterfeit the effect.  One can correct for that, but only if the neutrino mass hierarchy is known. 

As in the quark sector, the experimental verification and detailed
study of CP-invariance violation will require significant resources,
ingenuity, and patience. We recommend a program to resolve the question.  In general terms, sorting out the three unknowns of neutrino mixing, namely $\thb$, $\delta$, and the mass hierarchy, can be accomplished with a combination of:
\begin{enumerate} 
\item Long-baseline accelerator experiments in which sufficient matter is present in the beam path to provide sensitivity to the mass hierarchy via the effect of  matter on neutrino oscillations.  
\item Long-baseline accelerator experiments in which flavor conversion develops through the action of all three mixing angles, a prerequisite for observing CP violation.
\item Medium-baseline (a few km) experiments with reactors or accelerators to determine the magnitude of $\thb$ independent of the influence of CP violation and the mass hierarchy.
\end{enumerate}

\subsection{Neutrinos and the Unexpected}

Neutrinos may have properties beyond even
our new paradigm.  Such properties would again force a profound
revision in our thinking.  

\vspace{0.2 in}\vbox{

{\bf \noindent $\bullet$  \textcolor{red}{Are there `sterile' neutrinos? }} 

\begin{flushleft}
\frame{\hspace{0.1 in}
\begin{minipage}[]{2.8 in}
\begin{centering}
\vspace{0.1 in}
{\bf The Small, Silent Type}

{\it Neutrinos interact with other particles through the quiet
  language of the weak force.  This makes them elegant probes for new
  physics, because their voice is uncluttered by exchanges via the
  strong and electromagnetic interaction, unlike the gregarious quarks
  and charged leptons.  But the neutrinos that speak to us through the
  weak interactions may be accompanied by companions who are even
  quieter.  There are indications from experiments that these faint
  partners may exist.  }\vspace{0.1 in}
\end{centering}
\end{minipage}\hspace{0.1 in}
}

\end{flushleft}
}
 Elegant experiments at the world's largest electron-positron
  collider indicate that there are three and only three light
  neutrinos that interact with matter.  Other neutral fermions,
  lacking the universal weak interaction that characterizes the known
  neutrinos, would evade the inventory of species made in collider
  experiments.
  
   The speculated light neutral fermions capable of mixing with neutrinos
are known as `sterile neutrinos,' while the electron-flavor, the
muon-flavor and the tau-flavor neutrinos are referred to as `active neutrinos.' The existence of sterile neutrinos
mixed with the active neutrinos would affect the evolution of the universe and have important astrophysical consequences, in addition to its importance in the fundamental physics of particles. 
   
  Experimental studies of muon-flavor antineutrinos
  produced by antimuon decay at the Liquid Scintillator Neutrino
  Detector (LSND) at the Los Alamos National Laboratory,
  combined with the rest of existing neutrino data, hint at the
  possibility that sterile neutrinos may exist. If this is indeed the case, a more appropriate description 
of ``neutrinos'' may be best represented by something like 
Fig. \ref{fig:masses2}.
\begin{figure}[t]
\begin{center}
\includegraphics[height=3in]{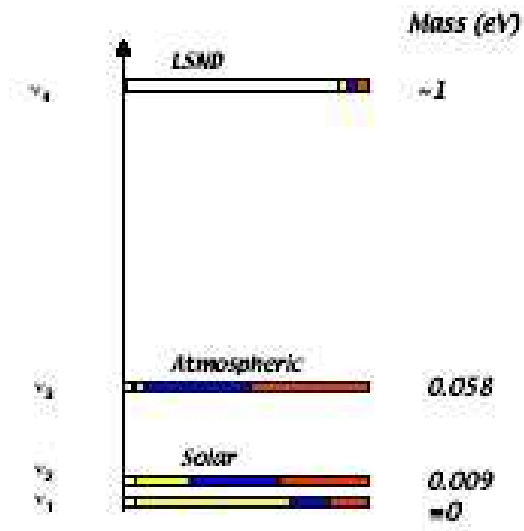}
\caption{A possible arrangement (hierarchy) of the masses of the neutrinos,
  based on oscillation data, with the additional input of evidence of
  a fourth type of neutrino from LSND.}.
\label{fig:masses2}
\end{center}
\end{figure}

In light of the importance of the physics implications, it is a priority to provide independent experimental confirmation.    This is the purpose of the MiniBooNE
experiment, currently running at Fermilab.

  \vspace{0.2 in}\vbox{
{\bf \noindent $\bullet$  \textcolor{red}{Do 
neutrinos have exotic properties?}}
 
\begin{flushleft}
\frame{\hspace{0.1 in}
\begin{minipage}[]{2.8 in}
\begin{centering}
\vspace{0.1 in}
{\bf A Still Closer Look}

{\it To our surprise, we have found that neutrinos have complex
  properties.  This hard-won discovery  ended the 70-year old picture of neutrinos as simple, massless
  objects.  Neutrinos have mass --- do they have other properties, too?  How do we find out more?  We must look carefully, as Ray Davis did years ago, to see if what we observe is always what we expect.    }
 \vspace{0.1 in}
\end{centering}
\end{minipage}\hspace{0.1 
in}
}
\end{flushleft}
}
 A wide range of exotic properties are possible in the neutrino
  sector.  These include magnetic and electric dipole moments,
  unexpected neutrino decays, and even violation of our most sacred
  fundamental symmetries.  We would be remiss not to search for
  these, since neutrinos have a long history of surprising us with
  their bizarre behavior.
 
Despite being electrically neutral, neutrinos may
have distributions of charge and magnetism
called electric and magnetic dipole moments.  This can only happen
with massive particles.  In the New Standard Model, the neutrino
magnetic moment is expected to be tiny, at least eight orders of
magnitude away from current experimental bounds. Reactor and
accelerator experiments in the next 10 to 15 years hope to improve the
sensitivity to neutrino magnetic moments by two orders of magnitude.
The observation of a nonzero effect would indicate the existence of
nonstandard physical effects mediated by new
particles at or above the electroweak symmetry breaking scale (about 100 GeV).

Massive particles may decay to lighter particles, so it is
theoretically possible for neutrinos to decay.  In the New Standard
Model, neutrinos decay to even lighter neutrinos and/or photons, and
the lifetime is expected to be absurdly long:
$\tau_{\nu}>10^{38}$~years.  Despite this, we should still search for
much shorter neutrino lifetimes, because that would be evidence that
our new paradigm is wrong.  Stringent bounds have been set for
neutrino decay into photons -- longer than billions of years. But
bounds on neutrinos decaying into new exotic matter are surprisingly weak.

There are many other deep physics principles that can be tested
through neutrino studies.  The discovery of effects such as the
violation of Lorentz invariance, of the equivalence principle, or of
CPT-invariance, to name only a few, would force us to redefine the
basic tools -- relativity, quantum mechanics -- we use in order to
describe Nature.  Physics and astrophysics would be led to the very
challenging but rewarding path of fundamental revision.  
 
 \vspace{0.2 in}\vbox{
{\bf \noindent $\bullet$  \textcolor{red}{What do 
neutrinos have to tell us about the
intriguing proposals for new models of fundamental physics?}}
 
\begin{flushleft}
\frame{\hspace{0.1 in}
\begin{minipage}[]{2.8 in}
\begin{centering}
\vspace{0.1 in}
{\bf Journey to a Grand Unified Theory}

{\it Like paleontologists, who must infer  the behavior of dinosaurs
  from a few remaining bones and fossils, physicists must reconstruct
  the behavior of particles at the high energies of the Big Bang from
  the clues provided by the low energy interactions we produce in the
  laboratory today.  Our recent new discoveries of the properties of
  neutrinos belong in the skeleton of a larger
  ``Grand Unified'' theory.  It is a strange looking beast, and further
  experimentation will be required before we can understand its full
  form.  }\vspace{0.1 in}
\end{centering}
\end{minipage}\hspace{0.1 in}
}
\end{flushleft}
}
 
  The discoveries about neutrinos have forced us to revise our robust
  and durable theory of physics, the Standard Model.  Until the
  question of whether neutrinos are their own antiparticles is sorted
  out, a clear path to the New Standard Model cannot be seen.  There
  are other tantalizing hints for physics beyond the New Standard
  Model. The ``running coupling constants'' seem to unify at some
  very large energy scale, leading to the strong belief that Nature
  can be described in terms of a simpler grand unified theory (GUT)
  that manifests itself as the Standard Model at lower, more
  accessible energies.

Neutrinos may turn out to play a major role in improving our
understanding of GUTs.  Some GUTs provide all the elements required to
understand small neutrino masses --- if they are their own antiparticles
--- and the matter-antimatter asymmetry of the universe via
leptogenesis. GUTs also provide relations among the quark mixing
matrix, the lepton mixing matrix, the quark masses, and the lepton
masses, in such a way that detailed, precise studies of the leptonic
mixing angles and the neutrino mass hierarchy teach us about the
nature of GUTs. In particular, the large mixing angles of the leptonic
mixing matrix provide an interesting challenge for GUTs.  The study of
neutrino masses and mixing provides a privileged window into Nature at
a much more fundamental level.

\subsection{Neutrinos and the Cosmos}

In the last few years the evidence for cold dark matter and dark energy in the cosmos have brought us face to face with the uncomfortable fact that we have no idea what 90\% of the universe is made of.   Neutrinos, oddly, {\it are} a component of dark matter, but a minor ingredient by mass.  Exactly how much, we do not know yet. On the other hand, despite being at a chilly 2K today, they were ``hot'' until the cosmos was billions of years old.  They may have played a  role in the formation of the vast skeins of galaxies in superclusters throughout the universe.     
 
\vspace{0.2 in}\vbox{ 
{\bf \noindent $\bullet$  \textcolor{red}{What is the role of neutrinos in
shaping the universe?}}
 
\begin{flushleft}
\frame{\hspace{0.1 in}
\begin{minipage}[]{2.8 in}
\begin{centering}
\vspace{0.1 in}
{\bf The First Neutrinos}

{\it Neutrinos were created in the cauldron of the Big Bang.  They orchestrated the composition of the first  nuclear matter in the universe. Their total mass
  outweighs the stars.  They played a role in the framing of the
  gossamer strands of galaxies.  We see in them the imprint of the
  cool matrix of neutrinos that fills and shapes the universe.
}\vspace{0.1 in}
\end{centering}
\end{minipage}\hspace{0.1 in}
}
\end{flushleft}
}
  
  The development of structure in the universe is determined by its
  constituents and their abundances. Neutrinos, due to their tiny
  masses, have streamed freely away from developing aggregations of
  matter until quite recently (in cosmological terms), when they
  finally cooled and their average speeds have decreased to
  significantly less than the speed of light. What is their role in
  shaping the universe? The answer to this question will not be known
  until the neutrino masses are known.

A stringent but model-dependent upper bound on the neutrino mass is
provided by a combination of neutrino oscillation experiments,
detailed studies of the cosmic microwave background radiation, and
``full sky'' galaxy surveys that measure the amount of structure in
the observed universe at very large scales.  It is a testament to the
precision of current cosmological theory that the fraction of the
universe's density contributed by neutrinos is only 5\% or less in
this analysis.  Laboratory measurements currently bound this number
from above at 18\%, and atmospheric neutrino oscillations set a lower
limit of 0.2\%.  A unique test of our current understanding of the
history of the universe will come from new experiments that directly
determine the neutrino mass.

Several experimental probes of astrophysics and cosmology will help
build a coherent picture of the universe at the largest
scales, including
\begin{enumerate}
\item Precision studies of the spectrum of the cosmic microwave 
background radiation;
\item Galaxy surveys;
\item Studies of gravitational weak lensing effects at extragalactic
  scales;
\item Precision determination of the primordial abundance of light
  elements; and
\item Studies of the nature of dark energy, such as surveys of 
distant type-Ia supernovae.
\end{enumerate}

\vspace{0.2 in}\vbox{ {\bf \noindent $\bullet$ \textcolor{red}{Are neutrinos the key to the
    understanding of the matter-antimatter asymmetry of the universe?}}
 \begin{flushleft}
\frame{\hspace{0.1 in}
\begin{minipage}[]{2.8 in}
\begin{centering}
\vspace{0.1 in}
{\bf  Neutrinos Matter}

{\it The universe is filled with matter and not antimatter.  But why?  In the initial fireball of the Big Bang, equal amounts of matter and antimatter were surely created. What gave the slight edge to matter in the race for total annihilation?  Surprisingly
  heavy members of the neutrino family could explain this
  asymmetry.  The light neutrinos we see today, the descendants of the
  heavy family, may hold the archaeological key.  }\vspace{0.1 in}
\end{centering}
\end{minipage}\hspace{0.1 
in}
}
\end{flushleft}}
   It is intriguing that lepton number and CP-invariance violation in
  the neutrino sector may be the answer to one of the most basic
  questions -- why does the universe we
  have observed so far contain much more matter than
  antimatter? In more detail, we would like to understand the
  following issue: in the distant past, the universe is very well
  described by a gas of ultra-relativistic matter and force
  carriers in thermal equilibrium. This thermal bath contained a very
  tiny asymmetry, around one extra proton or
  neutron, or `baryons,' for every $10^{10}$ baryons and
  antibaryons.
   
  As the universe cooled, almost all matter and antimatter
annihilated into light, and the tiny left-over matter makes up
all of the observable universe.  It is widely believed that the fact
that the primordial asymmetry was so small indicates that in even
earlier times the universe was described by a symmetric gas of
matter and antimatter, and that the asymmetry arose as the
universe evolved.  This dynamical generation of a matter-antimatter
asymmetry is referred to as `baryogenesis,' and the ingredients it
must contain were identified long ago: violation of C-invariance --
invariance of nature when particles are replaced by antiparticles --
and CP-invariance -- equivalent to time-reversal invariance;
baryon-number violation; and a time when the early universe was out of thermal equilibrium.  More than
just a matter of taste, baryogenesis is required in almost all models
for the universe that contain inflation, as the inflationary state of
the universe erases any finely-tuned matter-antimatter asymmetry one
could have postulated as present since the beginning of time.
Without baryogenesis, inflationary models predict a very boring,
matter-antimatter symmetric universe.

In the Standard Model with massless neutrinos, it is not possible to
generate the matter-antimatter asymmetry of the universe dynamically,
for a few reasons, including: (i) the CP-invariance violation present
in the quark sector is insufficient to generate a large enough baryon
asymmetry; and (ii) there are no physical processes that occur
significantly out of thermal equilibrium in a Standard Model gas,
at very high temperatures.  We only learned this recently, when
it became clear that the Higgs boson is not light enough.

Neutrino masses may come to the rescue. Not only do they provide new
sources of CP-invariance violation, they also provide new mechanisms
for generating the matter-antimatter asymmetry of the universe.  The
most popular mechanism for generating the matter antimatter-asymmetry
of the universe with the help of neutrino masses is called
`leptogenesis.' What is remarkable about several realizations of
leptogenesis is  that they relate the observed
matter-antimatter asymmetry of the universe to combinations of
neutrino masses, mixing angles and other free parameters. Hence, we may learn, by performing low-energy experiments, about whether
neutrino masses and mixing have something to do with the fact that the
universe is made of matter.

This is no simple matter, so to speak, and it may turn out that one can
never conclusively learn whether the answer is `yes' or `no.'  The main
reason is that it is likely that
the baryon asymmetry depends on parameters that describe Nature at
energies as high as $10^{15}$~GeV, an energy we simply
cannot access by direct experiment. Under these circumstances,
low-energy experiments can only probe particular combinations of the
leptogenesis parameters, and these may end up severely
underconstrained.

We have a plan for attacking this difficult problem.  First, we must
determine whether CP-invariance is violated in the leptonic sector.
Second, we must learn whether neutrinos are their own antiparticles,
and determine as well as possible the overall scale of neutrino
masses. It may turn out, then, that several realizations of
leptogenesis will be ruled out, or, perhaps, some very simple model
may fit all data particularly well. Further help may be provided by
non-neutrino experiments, including probes of the physics
responsible for electroweak symmetry breaking (is there low-energy
supersymmetry?, etc.) and searches for charged-lepton flavor violating
processes like $\mu\to e\gamma$. At that point, even if one cannot
{\it prove} whether leptogenesis is responsible for the matter-antimatter
asymmetry, we should have enough circumstantial evidence to believe it
or to reject it.
  
  \vspace{0.15 in}
  \vbox{ {\bf \noindent $\bullet$ \textcolor{red}{What can neutrinos disclose about
    the deep interior of astrophysical objects,  and about the mysterious sources of very high energy cosmic
    rays?}}
  \begin{flushleft}
\frame{\hspace{0.1 in}
\begin{minipage}[]{2.8 in}
\begin{centering}
\vspace{0.1 in}
{\bf Neutrino Odyssey}
  
{\it While the main focus of this story is on the physics of neutrinos
  themselves, it must not be forgotten that neutrinos can be used to
  probe both inner structure and outer limits.  They are messengers
  that come from deep in the heart of exploding stars and cataclysmic
  centers of galactic nuclei.    Through observation of these neutrinos,
  the fields of astrophysics and neutrino physics have illuminated each other in the past and will
  continue to  in the future.  }\vspace{0.1 in}
\end{centering}
\end{minipage}\hspace{0.1 in}
}
\end{flushleft}
}
   Neutrinos are the ultimate probe of astrophysical objects and
  phenomena. Neutrinos are deeply penetrating. Observing astrophysical
  neutrinos is the only way to look at the interiors of objects like
  the sun or the earth, and provides the only means of obtaining
  detailed information about the cataclysmic death of large stars
  in supernova explosions.  

Solar neutrino experiments over the past 30 years have, with our new understanding of the properties of neutrinos, provided convincing reassurance of our understanding of the sun.  However, we lack detailed confirmation of many important aspects.  The low-energy neutrino spectrum representing more than 99\% of the flux has been quantified only in radiochemical experiments that provide no detail of its structure.   The sun burns hydrogen to helium through two major cycles, and we have essentially no information about the one involving carbon, nitrogen and oxygen, other than that it is relatively weak.


Neutrinos may provide a means for understanding how the highest energy cosmic rays are produced and transported.  Unlike  protons, which,
along with heavier nuclei, are bent around by galactic and
extra-galactic magnetic fields, and photons, which are scattered by cosmic radiation
backgrounds, neutrinos  travel straight
to us, undeflected and unabsorbed.  Several probes of astrophysical
neutrinos are being built, developed, and studied, including:
\begin{enumerate}
\item Under-ice and underwater kilometer-size detectors of very high
  energy neutrinos, such as IceCube, in Antarctica;
\item Multi-kilometer-scale cosmic ray detector arrays, like the Auger 
experiment in Argentina;
\item New experiments to study the spectrum of neutrinos from the sun;
\item New detectors sensitive to coherent radio 
\phantom{hhhhhhhhhhhhhhhhhhhhhhhhhhhhhhhhhh}\newline
 \mbox{\ \ \ \ \includegraphics[width=5.5in]{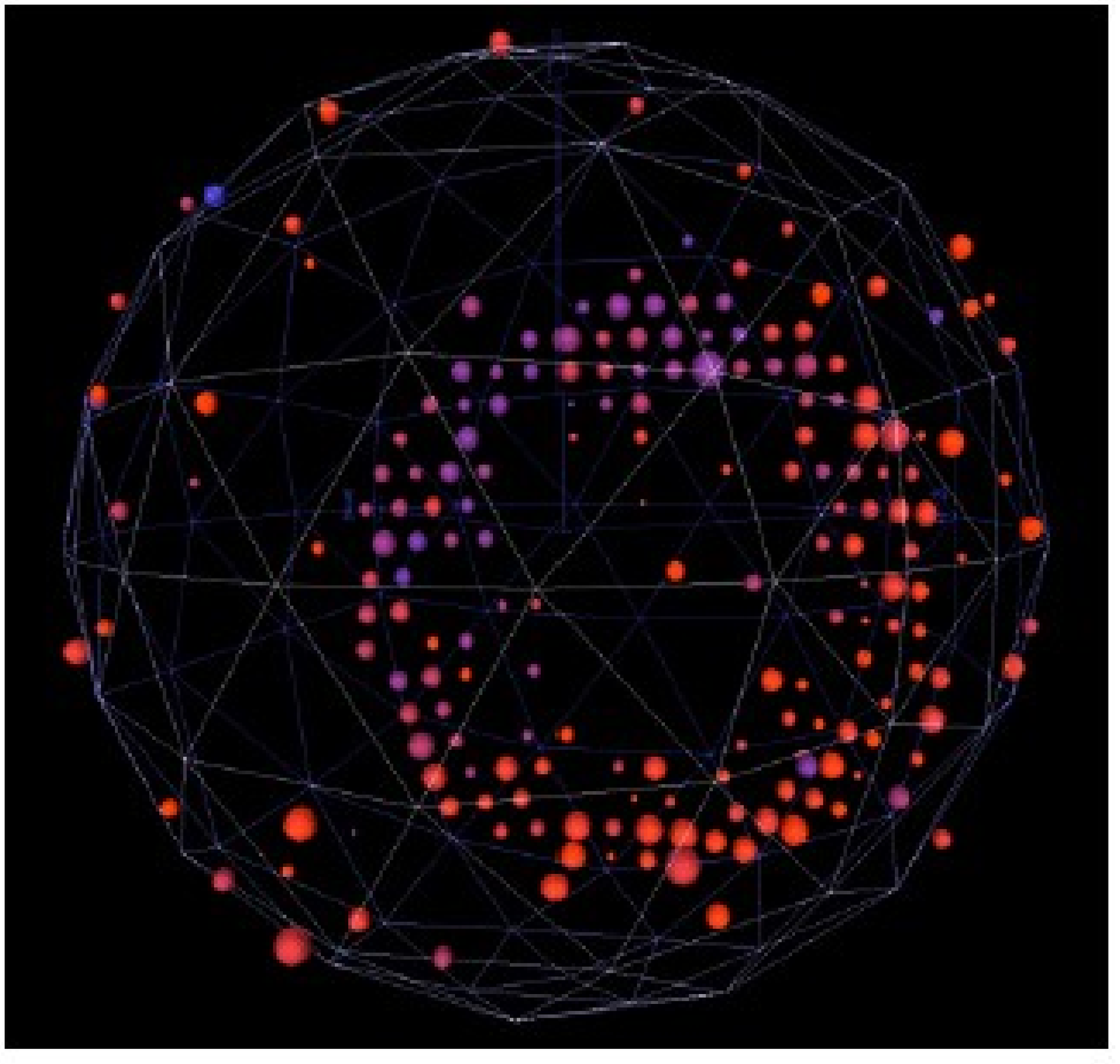}}
and acoustic waves 
produced by neutrino--matter interactions at extremely high energies, 
above $10^{15}$~eV, like the RICE and ANITA experiments; and
\item Efforts to observe galactic supernova explosions and the
  supernova neutrino background, expected to permeate space as a
  witness to all supernova explosions of the past.
\end{enumerate}

\begin{centering}
\begin{flushright}
{\em Below: A neutrino interaction in the MiniBooNE detector.  A muon moving faster than the speed of light in oil has been produced, and a  ring of Cherenkov light like an optical sonic boom is detected in photomultipliers mounted on the geodesic.  The size of each little sphere indicates the amount of light. }
\end{flushright}
\end{centering}

\clearpage

  \section {Current Program and International Context}
  
  The astonishing discoveries in neutrinos over the last decade
  promise to revolutionize our understanding of nature at the most
  fundamental level.  These discoveries have resulted from a broad
  range of experiments, many of  which were originally justified for
  different purposes. Some of these experiments continue, along with
  other new experiments that have been designed to provide yet more
  precise study of neutrino properties and perhaps offer even more
  revolutionary discoveries.

\begin{table*}[t!]
\begin{center}
\begin{tabular}{|c|c|}
\hline
AMANDA & Belgium, Germany, Japan, Netherlands, \\
& Sweden, United Kingdom, United States, Venezuela \\ \hline
CUORICINO & Italy, Netherlands, Spain, United States \\ \hline
KamLAND & China, Japan, United States \\ \hline
MiniBooNE & United States \\ \hline
SNO & Canada, United Kingdom, United States  \\ \hline 
Super-Kamiokande & Japan, Korea, Poland, United States  \\  \hline
 K2K & Canada, France, Italy,  Japan, Korea, Poland, \\ 
 & Russia, Spain, Switzerland, United States \\ \hline \hline
ANITA & United States \\ \hline
Auger & Argentina, Armenia, Australia, Bolivia, Brazil, \\
&  the Czech Republic, France, Germany, Italy, Mexico, \\
& Poland, Slovenia, Spain, United Kingdom, 
United States, Vietnam \\ \hline
Borexino & Belgium, Germany, Hungary, Italy, \\
&Poland, Russia, United States \\ \hline
IceCube & Belgium, Germany, Japan, Netherlands, New Zealand, \\
&Sweden, United Kingdom, United States, Venezuela \\ \hline
KATRIN & Czech Republic, Germany, Russia, United Kingdom, United States \\ \hline
MINOS & Brazil, France, Greece, Russia, United Kingdom, United States \\ \hline
RICE & United States \\ \hline \hline
\end{tabular}
\caption{Countries collaborating with the U.S. in our current and near-future experiments.}
\label{tab:whonu}
\end{center}
\end{table*}

  Neutrino physics enjoys a strong partnership between theorists and
  experimentalists, a relationship that drives the field forward.  The
  cross-cultural nature of the topic brings fresh ideas from
  astrophysics, cosmology, particle physics, and nuclear physics.
  International collaboration (see Table ~\ref{tab:whonu}) and
  competition have led to a healthy exchange of fresh ideas.  The range
  of experiments, in size and years of running, has allowed for both
  in-depth study and quick turnaround in investigating anomalies.
  Neutrino physics covers a broad range of experimental techniques and
  needs, and the existing program is already strong and rich in promise
  of new discovery. It is critical that, while future initiatives are
  undertaken, the current experimental programs be exploited as fully
  as possible. Furthermore, it is essential that the future program
  take account of the existing domestic and international efforts
  which are either already under way or planned for the next several
  years. With full use of the existing program, the future program
  outlined in this report has great potential for exciting new
  discoveries, even beyond the presently defined questions.

  \begin{figure*}[t!]
\begin{center}
\mbox{\ \ \ \ \ \ \ 
\includegraphics[width=6.5in]{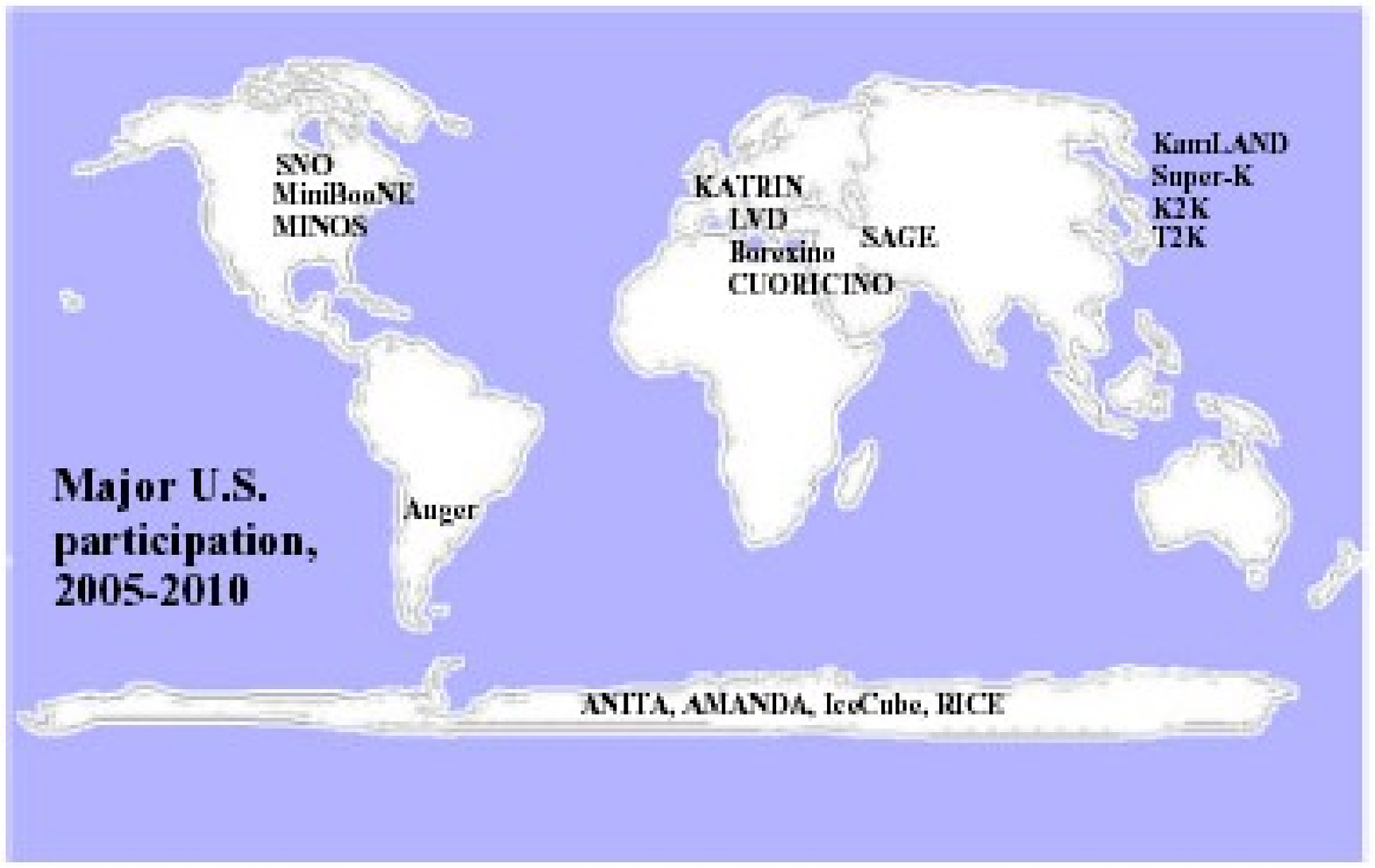}}
\caption{Neutrino experiments around the world.  The ones shown have 
  significant US involvement.}.
\label{fig:worldwide}
\end{center}
\end{figure*}

The existing U.S. experimental program (Fig. \ref{fig:worldwide}) is
in the process of addressing a substantial fraction of the important
topics we have just described.  It is critical that we provide strong
support to the current efforts, and where possible provide modest
additional investment in order to realize the best return from these efforts. Some of the important ongoing experiments either in the
 {U.S.} or with substantial  {U.S.} participation are:
\begin {itemize}
\item {\bf The UHE Program:} The U.S. has played a major role in the development of methods for the detection of ultra high energy cosmic rays.  AMANDA has pioneered the use of the Antarctic ice
   as a neutrino telescope. It is currently taking data and
  will be integrating with the km$^3$-sized IceCube over the next
  year or so.  Radio and acoustic methods have been explored with GLUE, FORTE, and SAUND, and the program continues with RICE and ANITA in the Antarctic.  U.S. scientists also collaborate in large lake and ocean cosmic-ray detectors, Baikal, ANTARES, NEMO,  and NESTOR.
  
\item {\bf KamLAND:} Recent results from the KamLAND experiment,
  located in Japan, show a clear energy dependent oscillation effect
  that not only clearly agrees and confirms solar neutrino
  oscillations but also strongly constrains the possible range of
  $\Delta m_{12}^2$.  Kamland continues to collect data and we
  anticipate that the final results will provide a precision
  measurement of this parameter for which we do not expect any
  improvement for the foreseeable future.
  
\item {\bf MiniBooNE:} This U.S.-based experiment is running in
  neutrino mode, and benefiting from continuous improvements in
  the Fermilab Booster delivery of beam.  Should the LSND $\bar{\nu}_{\mu}$ to $\bar{\nu}_e$ transition signal be confirmed, the collaboration plans
  additional experiments, described in the superbeams working group
  report.  As discussed in {\em Recommendations}, a decisive resolution of this question is essential, which may require additional studies with beams of antineutrinos. 
  
\item {\bf SNO:} The SNO experiment, in Canada, has provided
  crucial experimental evidence contributing to the proof that the
  solar neutrino deficit results from flavor transitions from $\nu_e$
  to some combination of $\nu_\mu$ and $\nu_\tau$.  To complete its physics program, SNO is now
  preparing the detector for operations with $^3$He neutron counters
  in order to improve sensitivity to the mixing angles $\tha$ and
  $\thb$.

\item {\bf Super-Kamiokande and K2K:} Decisive evidence of
  oscillations in atmospheric neutrinos has come from
  Super-Kamiokande, and the oscillation phenomenon is now also seen in
  K2K with neutrinos from the KEK accelerator.  These experiments,
  located in Japan, are impressive for the breadth and quality of
  results on atmospheric, accelerator, and solar neutrinos.
  Super-Kamiokande is currently operating with about half its full
  photomultiplier  complement, and will undergo refurbishment to the
  full coverage in 2005.

\end {itemize}

Recognizing the importance of neutrino studies, the  {U.S.} is
already committed to several new experiments that are well into the
construction phase:
\begin {itemize}
  
\item {\bf ANITA:} This balloon-borne radio telescope, to be launched
  in the Antarctic, is designed to detect very high energy neutrinos
  resulting from the GZK effect.  A characteristic pulse of
  radio energy is produced by  the
  intense shower of particles when such neutrinos interact in the ice.
  ANITA is expected to provide the first
  sensitivity to these putative neutrinos.
  
\item {\bf Auger:} Auger is a 3000-km$^2$ air shower array currently
  under construction in Argentina with substantial  {U.S.} involvement.
  Auger's primary goal is the study of very high energy air
  showers, including those produced by neutrinos at and above the
  GZK cutoff.

\item {\bf Borexino:} This experiment,  at the Gran Sasso Laboratory in Italy, is aimed at a
  measurement of solar neutrinos with energy spectrum sensitivity and
  ability to measure the flux from $^7$Be decays.  Construction is essentially complete, but operations have
  been delayed. It is hoped that operations can begin in 2005. As we discuss in {\em Recommendations},  a prerequisite in physics with solar neutrinos is a determination of the  $^7$Be neutrino flux to an accuracy of 5\% or better.
    
\item {\bf IceCube:} This is a km$^3$ high-energy neutrino
  observatory being built in the ice cap at the South Pole. It is an
  international collaboration with primary support coming from the
  NSF. It will very substantially extend sensitivity to possible
  astrophysical point sources of neutrinos.

\item {\bf KamLAND Solar Neutrinos:} Plans are developing to upgrade the KamLAND detector in Japan  to permit a lower energy threshold in order to detect solar neutrinos from $^7$Be decay.  Both Japan and the  {U.S.} are participating.
  Because the measurement of $^7$Be neutrinos represents a substantial
  experimental challenge, it is likely that two independent
  experiments will be necessary  to reach the desired 5\% accuracy.
  
\item {\bf KATRIN:} The KATRIN experiment is under construction in
  Germany.  It involves an international collaboration focused on improving
  the sensitivity to direct neutrino mass measurement in tritium beta
  decay. KATRIN represents an excellent example of  {U.S.} groups 
  working together with international collaborators to build a single
  facility with unique capabilities.

\item {\bf MINOS:} The Fermilab NuMI beamline will be complete late in 2004 and
  MINOS beam operations will begin. This  {U.S.}-based experiment will
  offer precision measurements of oscillation parameters and extension
  in sensitivity to $\nu_e$ appearance.  The sensitivity of MINOS
  depends on the number of protons that can be delivered. As we discuss in {\em Recommendations}, continued improvements
  in the  proton intensity are necessary for the present Fermilab experiments to meet their physics goals.
    
\item {\bf RICE:} RICE, which seeks to observe
  neutrinos at the highest energies, has pioneered the use of an array of radio
  antennas on the surface of the Antarctic ice for the observation of energetic charged particles.  It is currently taking data.
  Theoretical estimates of neutrino fluxes suggest that substantially
  larger arrays may be required for positive observation of ultra-high
  energy neutrinos.

\end {itemize}

In addition to the existing or soon-to-exist experiments with
significant  {U.S.} involvement,  there are important new experiments being
planned or built abroad that will inform the planning for a future
 {U.S.} program. In discussing these future prospects, we do not include
all possible future activities but take some account of the relative
advancement of the proposal or status of construction.  Some of the major
experiments being planned/built, of which our proposed  {U.S.} program has
taken explicit account, are:
\begin {itemize}
  
\item {\bf CNGS:} Two experiments, ICARUS, and OPERA, are under
  construction at the Gran Sasso Laboratory in Italy for use with the
  CERN-Gran Sasso neutrino beam, which will start operation in 2006.
  These experiments will search for evidence of $\nu_\tau$ appearance
  and, along with MINOS, will extend the sensitivity to $\nu_e$
  appearance. CERN, located in Switzerland, is working to increase SPS
  proton intensity in order to maximize the physics output.
  
\item {\bf Indian Neutrino Observatory (INO):} A large magnetized
  atmospheric neutrino detector is being proposed for construction in
  India. This detector may provide sensitivity to the neutrino mass
  hierarchy.
  
\item {\bf LVD:} LVD is an 800-ton liquid scintillator detector at
  Gran Sasso Laboratory in Italy,  with
  sensitivity to a galactic supernova.
  
\item {\bf Mediterranean Neutrino Observatory:} There are three
  underwater neutrino telescopes currently under development in the
  Mediterranean, NESTOR, NEMO, and AN-TARES.  It is anticipated that
  these development projects will result in a final project to build a
  single km$^3$-size detector. This will add a northern complement to
  the IceCube Detector. No  complementary {U.S.} project is proposed for the
  northern hemisphere, and modest  {U.S.} collaboration may develop
  on the effort in the Mediterranean.

\item {\bf Neutrinoless double beta decay:} There are many
   R\&D programs worldwide in double beta decay, some of which
  include operating experiments.  Among isotopes receiving the most attention are $^{76}$Ge, $^{100}$Mo, $^{130}$Te and $^{136}$Xe.   The
  NEMO III experiment in the Modane Laboratory in France is collecting
  data with kilogram quantities of several enriched isotopes, and
  features particle tracking for event identification.  Cuoricino is a
  calorimetric experiment operating with kilogram quantities of
  natural tellurium.  Both experiments plan expansions.  A controversial
  analysis of data from the Hei\-del\-berg-Mos\-cow experiment that used
  approximately 10 kg of enriched $^{76}$Ge yields evidence for an effective
  neutrino mass greater than 0.1 eV.

\item {\bf Reactor experiments:} The proposed
  Double CHOOZ experiment in France will use the existing underground
  space where the first CHOOZ experiment was performed, along with a
  near detector to reduce the systematic uncertainty.  A proposal has
  been submitted and is in the approval process. KASKA is an
  experiment being planned for the Kashiwazaki reactor site in Japan.
  Both of these experiments have a sensitivity goal of  $\sin^2
  2\theta_{13} \lt 0.03$ at 90\% CL for $\Delta m^2=0.002$ eV$^2$.  In addition, there are  U.S. initiatives for experiments aiming at the $\sin^2
  2\theta_{13} \lt 0.01$ level that would be carried out at reactor sites in Brazil, China, or the U.S.

\item {\bf SAGE:} The SAGE gallium experiment in Russia is unique in
  its sensitivity to neutrinos from the proton-proton ($pp$)
  interaction and $^{7}$Be decays in the sun.  With termination of the
  GNO gallium experiment at Gran Sasso, discussions have commenced
  about combining the SAGE and GNO collaborations. Formal
  participation by {U.S.} groups in SAGE has ended, but cooperation in
  this important experiment continues.

\item {\bf T2K:} T2K will use the new 50 GeV accelerator, starting in 2009 at Tokai,
along with the Super-Kamiokande Detector to improve sensitivity to $\nu_e$
appearance about a factor of 5--10 beyond MINOS and CNGS. Due to the 295-km baseline, T2K is almost insensitive to matter effects. This makes 
relatively cleaner measurements for $\theta_{13}$ and $\delta_{CP}$ but 
does not provide sensitivity to the mass 
hierarchy. For that reason, that type of experiment is a good complement to
a longer-baseline experiment with sensitivity to matter effects, such that
the combination of the two provides clean separation of all of the associated parameters. There is {U.S.} 
participation in T2K with developing plans on the scope of that 
participation. T2K is an important part of a coherent international effort
necessary to measure all of the important oscillation parameters.
\end {itemize}

Several related experimental programs provide crucial data for
better understanding results from the neutrino experiments. Some of
these include:
\begin {itemize}
\item {\bf Nuclear Physics Cross Sections:} Nuclear-physics cross
  section measurements, such as for the fusion of $^3$He with $^4$He and for the reactions of protons with  certain radioactive nuclides, will continue to be critical to understanding the sun and
  supernovae.  
\item {\bf Cosmic-Ray and Astrophysics Measurements:} Cosmic-ray and astrophysical measurements are
  important to  an understanding and
  prediction of observed neutrino sources.
\item {\bf Cosmology connections to neutrinos:} Measurements of the
  cosmic microwave background (CMB) and large scale structure continue
  to offer very interesting promise of placing limits on, or even
  observation of, an effect resulting from neutrino mass in the range
  of 0.1 eV.
\end {itemize}

A final consideration is support for a strong theory effort on the
broad set of issues in neutrino physics. Theoretical efforts in
neutrino physics have played a fundamental role in interpreting the
wide range of revolutionary experimental results and building a
coherent, yet still incomplete, picture of the new physics uncovered
by the discovery of neutrino flavor transitions.  Among the triumphs
of such efforts are computations of the solar neutrino flux,
development of the neutrino oscillation formalism including the
effects of neutrino propagation in matter, and determination of the
effects of neutrinos in Big-Bang nucleosynthesis, large scale
structure formation, and the distortions of the cosmic microwave
background radiation.

It is also part of the theoretical efforts to establish connections
between the new discoveries in neutrino physics and our most
fundamental understanding of matter,  energy, space, and time. Significant
advances have been made in several arenas, including establishing
connections between neutrino masses and leptonic mixing with the
concept of grand unification, establishing a relationship between
neutrino masses and the matter-antimatter asymmetry of the Universe
(through leptogenesis), and developing different predictive mechanisms
for understanding the origin of neutrino masses in a more satisfying
and relevant way.

Finally, due to the particular interdisciplinary nature of neutrino
physics, theory has played the absolutely essential role of
 integrating results and developments in astronomy, astrophysics,
cosmology, high-energy and low-energy particle physics, and nuclear
physics. As new discoveries arise in all of these disciplines,
theoretical guidance and integration will continue to be
indispensable.
 \vspace{0.2in}
\begin{centering}    
 \mbox{\includegraphics[width=6.5in]{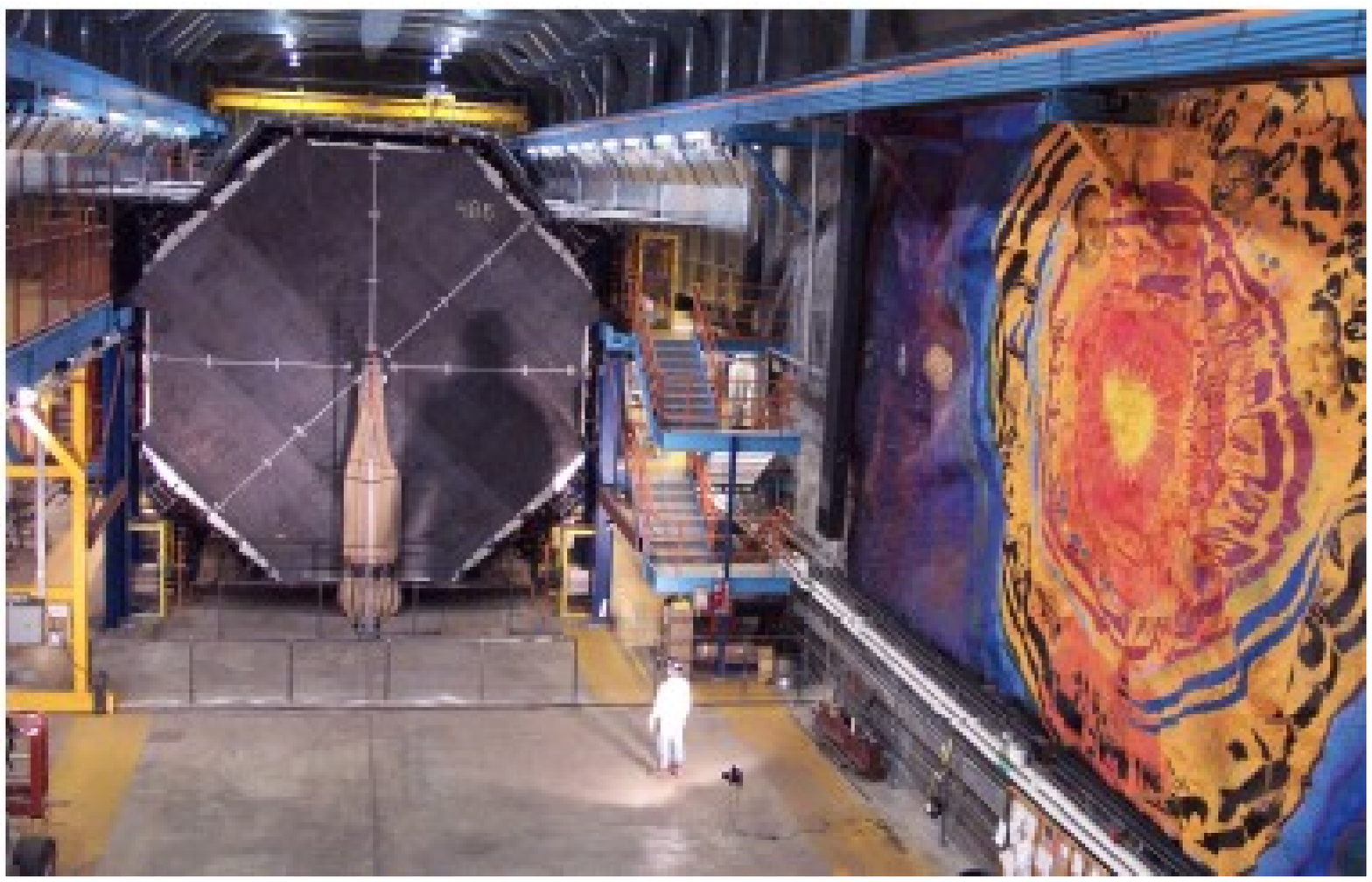}}
\end{centering}

 \begin{centering}
\begin{flushright}
\vspace{2.5 in}
{\em Below: The MINOS detector in the Soudan mine in Minnesota.  A huge mural brightens the work area. }
\end{flushright}
\end{centering}

\clearpage

\section{Recommendations}

Our recommendations for a strong future {U.S.} neutrino physics
program are predicated on fully capitalizing on our investments in the current program.  The present program includes the long\-est baseline neutrino beam and a high-flux short baseline beam, both sited in the U.S.\ \  Elsewhere, American scientists and support are contributing in important ways to the burgeoning world program in neutrino physics, including a long-base\-line reactor experiment in Japan, solar and atmospheric neutrino experiments in Canada,  Italy, Japan, and Russia,  a direct mass measurement in Germany, ultra high energy astrophysics experiments in Antarctica and Argentina, and other experiments.  We congratulate not only the scientists involved but also the Agencies for their perceptive support of this developing program, which has been so spectacularly  fruitful.  

Four  issues deserve special mention:
\begin{enumerate}
\item Support for continued increases of proton intensity for Fermilab
   neutrino experiments, as is necessary for the present experiments to meet
   their physics goals.
\item Support for decisive resolution of the high-$\Delta m^2$
   puzzle.  This issue  is currently addressed by a single experiment now running in a neutrino
   beam at Fermilab.  Ultimately, a decisive resolution of the puzzle may  require additional studies
   with beams of antineutrinos. 
\item Support for determination of the $^7$Be solar neutrino flux.  Such measurements  are currently in the program of two underground detectors, one in 
   Italy and the other in Japan. 
\item Continued support for enhanced R\&D focusing on new techniques for detecting
   neutrinos above $10^{15}$ eV from astrophysical \newline sources.   This capability would open
   a new window to astrophysics with significant discovery potential.
\end{enumerate}


Turning to the recommendations for the future, we preface our remarks by drawing attention to some basic elements in common:

\begin{enumerate}
\item In every instance the need for suitable underground detector facilities emerges.   A successful neutrino program depends on the availability of such underground space.
\item The precise determination of neutrino cross sections is an essential ingredient
in the interpretation of neutrino experiments and is, in addition, capable of
revealing exotic and unexpected phenomena, such as the existence of
a neutrino magnetic  dipole moment. Interpretation of atmospheric and long-baseline
accelerator-based neutrino experiments,
understanding the role of neutrinos in supernova explosions, and predicting the abundances of 
the elements produced in those explosions all require knowledge of neutrino cross
sections.  New facilities, such as the Spallation Neutron Source, and existing neutrino beams can be used to meet this essential need.
\item It is important that at least two detectors worldwide should be operational which, in addition to their other physics roles, are continuously sensitive to a galactic supernova.   
\end{enumerate}

Our recommendations have their genesis in central questions in neutrino physics: What are the masses of the neutrinos? How and why do they mix? Are neutrinos their own antiparticles? Is CP symmetry broken by neutrinos?   A comprehensive understanding of fundamental physics and of the universe rests upon the answers to such questions.


$\bullet$  {\bf We recommend, as a high priority, that a phased program of
     sensitive searches for neutrinoless nuclear double beta decay be
     initiated as soon as possible.}

   Neutrinoless double beta decay is the only practical way to discover
   if neutrinos are their own antiparticles and, thus, a new form of
   matter.  Without this information, the construction of the New Standard Model cannot be completed. The lifetime for neutrinoless double beta
   decay  is inversely proportional 
   to an effective neutrino mass. Hence, in order to observe a signal experimentally, not only must
   the neutrinos be their own antiparticles, they must also be sufficiently massive.

   We recommend a phased approach with successively larger detectors and lower backgrounds.  The first experiments should address masses of a few
   tenths of an eV.   This is the `degenerate' mass scale in which the three neutrino masses are nearly equal, and it is the range in which the large-scale structure of the universe would be affected.  From  cosmological  and  existing double beta decay data,  controversial arguments have been made that the neutrino mass is actually of this size. For this mass range, neutrinoless double beta
   decay can  be discovered and precisely measured with isotopic samples of
   approximately 200 kg in a period of 3 to 5 years.

If neutrinoless double beta decay is
   not observed in the 200-kg experiments, then a second phase of experimentation  with 1-ton isotopic samples should be initiated to search in the 20 to 55 meV mass range.  That is the range given by the observed atmospheric neutrino oscillation signal if the  mass hierarchy is non-degenerate and inverted.  A non-degenerate, {\em normal} mass hierarchy with effective masses below 20 meV requires sample sizes of hundreds of tons.   For that scale of experiment substantially more R\&D will be necessary.

  The issue is singularly important, the experiments are difficult, and there is, moreover, some uncertainty in the theory that applies to each candidate nucleus.  Hence it is prudent to pursue more than a single scalable technique with different isotopes and an expanded R\&D effort.  Worldwide, only four collaborations (two
   predominantly European and two predominantly {U.S.}) are likely to propose
   viable 200-kg  experiments (with $^{76}$Ge, $^{130}$Te, and $^{136}$Xe) in the near future.  It is conceivable that two of the groups will merge, leaving three efforts among which the {U.S.} will play a major role in two, and a secondary role
   in the third. 
   
   The {U.S.} is well positioned to make a significant
   contribution to this program.  However,  these experiments all require that  
   appropriate underground facilities at moderate to substantial depth be available.

\vspace{0.1in}

$\bullet$  {\bf We recommend, as a high priority, a comprehensive {U.S.}
    program to complete our understanding of neutrino mixing, to
    determine the character of the neutrino mass spectrum, and to
    search for CP violation among neutrinos.  This program should have
     the following components:
\begin{itemize}
\item An expeditiously deployed multi-detector reactor experiment with
   sensitivity to  \boldmath$\overline{\nu}_e$ disappearance down to 
\boldmath{$ \sin^22\thb
   = 0.01$}, an order of magnitude below present limits.
\item A timely accelerator experiment with comparable 
\boldmath$\sin^22\thb$ sensitivity and sensitivity to the 
mass-hierarchy \newline through matter effects.
\item A proton driver in the megawatt class or above and neutrino
   superbeam with an appropriate very large detector  capable of observing CP violation and measuring  the  neutrino mass-squared differences and mixing parameters  with high precision.
\end{itemize}
}

The discovery of neutrino oscillations has provided completely new
information about neutrino masses and mixing.  To complete our
understanding of mixing and the mass hierarchy, to discover whether or not  the CP symmetry is violated by neutrinos, 
and to be sensitive to unanticipated new physics,  a flexible program with several complementary experiments is
necessary.

Knowledge of the presently unknown value of the mixing angle $\thb$ is a 
key factor in all of these objectives.  Determination of this important
parameter, or at least a stringent limit on it down to $\sin^22\thb =
0.01$, can be established with a relatively modest scale reactor
experiment.  We strongly urge the initiation of a
reactor based multi-detector experiment with this sensitivity as soon
as possible.

A new long-baseline experiment using the existing NuMI beamline at Fermilab and a 
beam upgraded to 0.4 MW would be sensitive to combinations of
the mixing angles $\thb$ and $\thc$, the phase $\delta$, and the
mass-squared difference $\dmc$. Furthermore, if $\sin^22\theta_{13}$
is large enough, such 
an experiment in concert with other experiments can potentially determine the neutrino mass hierarchy through matter effects. 
Such an experiment should be roughly 10 times more 
sensitive to
$\nu_e$ appearance than the long baseline experiment currently under way at Fermilab
and, if done in a timely manner, would capitalize
on the considerable investment in NuMI. 

Given that the value of $\thb$ is presently unknown, should  the accelerator and reactor
experiments be done in sequence or contemporaneously?  
We  strongly recommend the contemporaneous strategy.  First,  accurate
determinations of  $\thc$, $\dmc$ and either a stringent upper
limit or a value for $\thb$ are of central importance to an understanding of the
origin of neutrino masses and mixing.  Second, in almost any conceivable scenario, it will be essential to have the complementary and/or confirmatory information from these different techniques.  Third, we draw attention to
the unique and time-sensitive opportunity for the {U.S.} to build a
strong accelerator-based neutrino physics program, with real discovery
potential, that will be a major contributor in the rapidly advancing
world program. 

Even without knowing the outcome of the initial steps in the program, 
it is clear that very
large-scale, long-baseline experiments will provide the best sensitivity to
all the oscillation parameters as well as to possible unanticipated new
physics. They also provide the only possibility for quantitatively exploring
CP-invariance violation in the neutrino sector. A 
proton driver in the
megawatt class or above used to produce a neutrino superbeam, together
with a detector of more than 100
   kilotons mass,  should be able to
probe all aspects of three-generation neutrino mixing, unambiguously
determine the mass hierarchy, and provide definitive information on the
amount of CP-invariance violation, as long as $\sin^22\thb$ is larger 
than about $ 0.01$.    If $\sin^22\thb$ is smaller still, a neutrino factory 
will be required, because of its potential freedom  from backgrounds.
Such a facility likewise requires an intense proton driver.
The intense proton driver and detector would each provide benefits across a wide spectrum of fundamental physics in addition to neutrino physics. 

Because of the long lead time in designing a new intense proton
driver, a decision  whether to embark on such a program should be
made as soon as practicable.  With their existing accelerator
infrastructures and capabilities, either Brookhaven or Fermilab would
be natural sites, and both laboratories have been working on designs.
A comprehensive study of the scientific, technical, cost, and
strategic issues will be necessary.

Massive detectors have been key to the recent 
revolution in
neutrino physics.  Their significant cost is more appropriately 
justified by the diverse physics
program made possible by a multipurpose detector.  Such a detector
should be capable of addressing problems in nucleon decay, solar
neutrinos, supernova neutrinos, and atmospheric neutrinos in addition to
long-baseline neutrino physics.
The broad range of  capabilities, however, can only be realized if it is
built deep enough underground. If such a detector  is to be sited in the {U.S.},
appropriate new underground facilities must be developed.  

A high-intensity neutrino factory or a `beta-beam' facility is the ultimate tool in neutrino
physics for the long term, and may be the only facility capable of definitively
addressing some of the physics issues.  Neutrino factories  and beta beams require, respectively, development of  a  muon storage ring or a radioactive-ion storage ring, which provides
intense, high energy muon and/or electron neutrino beams with well understood
energy spectra and very low background levels.   
Neutrino factories are presently the focus of the
{U.S.} development program, and there is a significant collaboration with
Europe and Japan.  The neutrino factory R\&D program needs increased levels of support if the facility is to be realized in the long term.

The overall program must be considered in an international context.
Reactor experiments less sensitive than the one recommended here are
being considered in France and Japan.  An interesting and extensive
off-axis superbeam program is under construction in Japan.  Like the 
recommended {U.S.} program, it is sensitive to a combination of 
parameters.  The programs are complementary because only the {U.S.} 
program has sufficiently long baselines to provide good sensitivity to the mass
hierarchy through matter enhancement. With both the {U.S.} and 
international programs, we may confidently anticipate a thorough 
understanding of neutrino mixing.

\vspace{0.1in}

$\bullet$ {\bf 
   We recommend the 
   development of a spectroscopic solar neutrino experiment capable
   of measuring the energy spectrum of neutrinos from the primary $pp$
   fusion process in the sun.}

The experiments that first established neutrino flavor transformation
exploited neutrinos from the Sun and neutrinos produced in the earth's
atmosphere.  These sources continue to be used in the present program
of neutrino experiments.  Natural neutrino
sources are an important component of a program seeking to better
understand the neutrino and at the same time aiming to use neutrinos
to better understand astrophysical sources.

A measurement of the solar neutrino flux due to $pp$ fusion, in
comparison with the existing precision measurements of the high\-er-energy
$^8$B neutrino flux, will demonstrate the transition between vacuum
and matter-dominated oscillations, known as the 
Mik\-heyev-Smirnov-Wolfenstein effect.  In combination with the 
essential
prerequisite experiments that will measure the $^7$Be solar neutrino
flux with an accuracy of 5\%, a measurement of the $pp$ solar
neutrino flux will allow a sensitive test of whether  the Sun
shines exclusively through the fusion of light elements.   Moreover, 
the neutrino luminosity of the Sun today is predictive of the Sun's 
surface temperature some 10,000 years in the future because 
neutrinos, unlike photons, travel directly from the center of the Sun 
to the earth.

Low-energy solar neutrino experiments need to be located in very deep 
underground sites in order to achieve the required reduced levels of
background. If one is to be located in the {U.S.}, adequate underground
facilities are required. 

\vspace{0.1in}

A coordinated program such as we recommend has enormous discovery potential, and builds naturally upon the successes already achieved in the U.S. program.  It is a rare and wonderful circumstance that the questions of fundamental science can be so clearly formulated and so directly addressed.  

   \section{Timeline and Branch Points}
   How will the program we have recommended here evolve with time,
   what are the branch points at which new information will illuminate
   the course ahead, and how do the {U.S.} and world programs move forward
   in mutual cooperation?  In Fig. \ref{fig:timeline}, a schematic
   timeline illustrates a feasible and appropriate schedule for the
   research.
\begin{figure*}[ht]
\begin{center} 
\phantom{hhhhhhh}\includegraphics[width=5in]{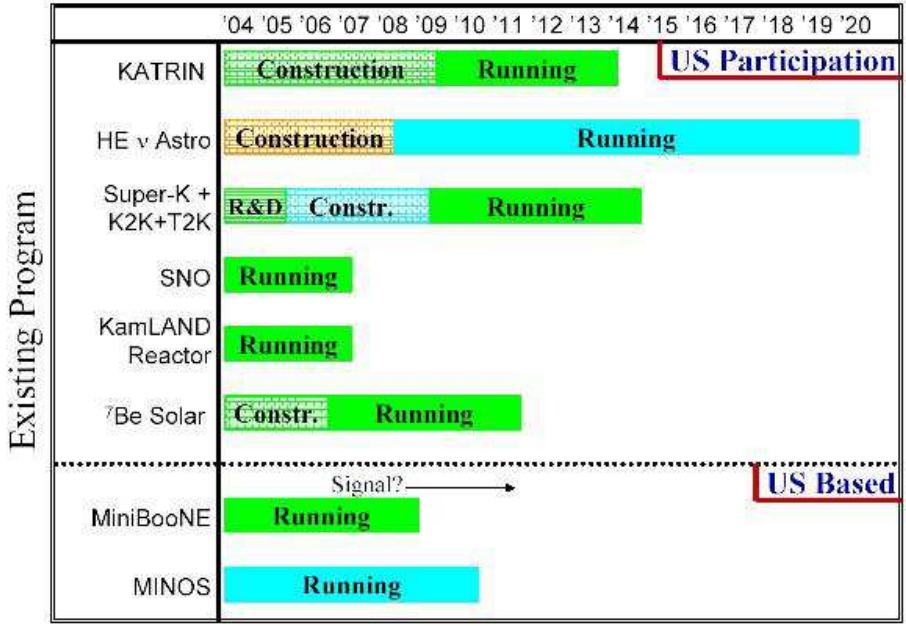}\newline
\phantom{hhhhhhh}\newline
\phantom{h}\includegraphics[width=5in]{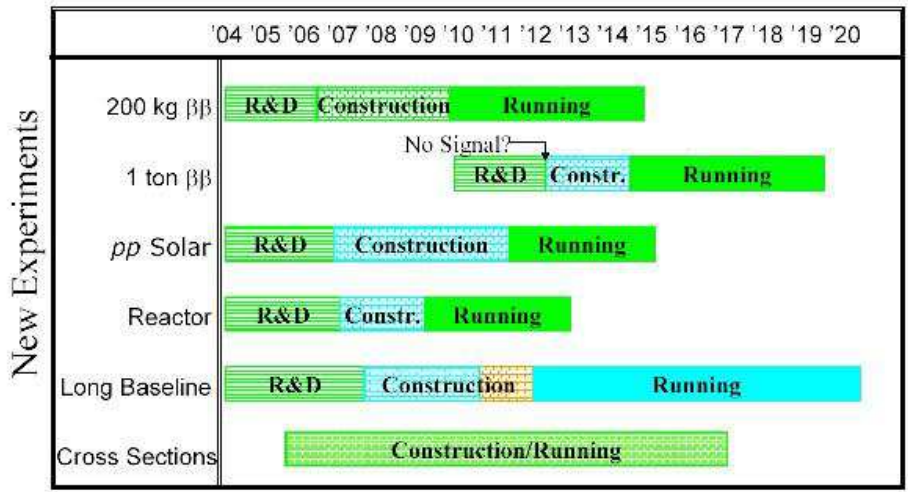}\newline
\phantom{h}\newline
\phantom{h}\includegraphics[width=5in]{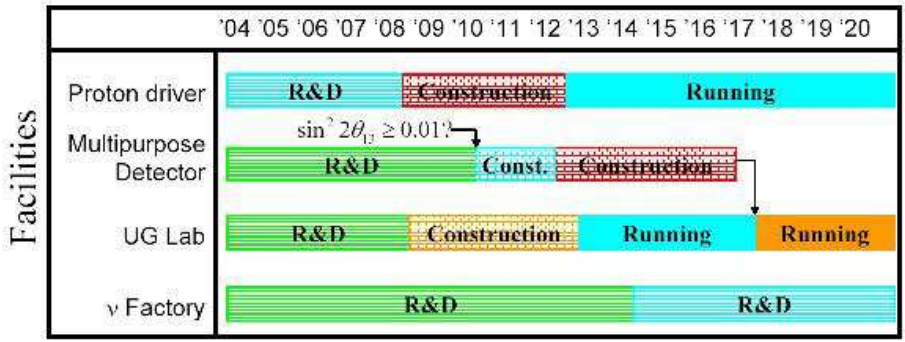}
\caption{An
  approximate indication of the development of our recommended
  neutrino program with time.  Some branchpoints are also indicated. Colors indicate U.S. contribution. Green: $\leq$ \$10 M per year. Blue: \$10
  -- 40 M per year. Orange: \$40 -- 100 M per year.  Red: $\geq \$100$M per
  year.  Barred: R\&D. Hatched: Design and construction.  Solid:
  operations.}.
\label{fig:timeline}
\end{center}
\end{figure*}
As information is gained, a number of experimental
programs have branch points.  It is difficult to predict all of the
possible future branches.  Here we note those that are clearly discernible.

The neutrinoless double beta decay program will reach a decision point
after the results of the 200 kg experiments are known.  In the event
that no signal is seen, the likely branch is to  larger detectors
sensitive to the `atmospheric mass' range.  A positive signal at any stage will require experiments with other isotopes to confirm such a fundamental scientific observation and to reduce the influence of theoretical uncertainties in the quantitative result for the effective neutrino mass; because the experiments take many years, it is necessary to initiate more than one at each branch.  

The direction that the comprehensive program of oscillation parameter
measurements takes in the future depends on the value of the parameter
$\sin^2 2\theta_{13}$.  If this parameter is larger than 0.01, the
program we have outlined  will accurately
determine some of the underlying physics, while the recommended proton
driver and very large detector will be necessary for a quantitative understanding of the extent of CP violation among the neutrinos.   If, on the other hand, this parameter is less than 0.01,
information on  neutrino mixing  will be provided by the
proton driver and appropriate very large detector, but the search for CP violation must await the neutrino factory.

The resolution of the LSND question also represents an important
branch point, although in this case, observation of a signal would
call for augmentation of the program presented in this document. The
current program would continue as presented, but with additional goals
and accompanied by a suite of appropriate new experiments to further explore this new
phenomenon.  

\section{Conclusions of the Study}
   
   In this study, neutrino physicists, accelerator physicists, and astrophysicists have worked together to identify the most
   exciting scientific opportunities for the future of neutrino
   physics.  We have prioritized these needs, dividing our findings
   into two high priority recommendations that we concluded are crucial
   for the continued advancement of the field, and one
   that  would substantially enhance the {U.S.}
   program through its added discovery potential.  They  represent but a small subset of
   the interesting ideas that emerged from the study, ideas
   reported in the appendix of Working Group Reports.  This collection, which we believe represents the   future in each study area,
   underlines the intellectual richness of the field.
   
   Out of this activity has emerged a program for which the whole
   will be greater than the sum of its parts.   The program is coordinated to maximize
   results and minimize duplication, taking into account the worldwide program.  Our recommendations encourage international
   cooperation, in order to leverage {U.S.} investment.  Our choices are
   interdisciplinary, exploiting the excitement of connecting results
   from wide-ranging disciplines.   Just as the science represents the convergence of many disciplines, so too will the continued support of many Agency Divisions and Offices be needed to bring  it to fruition.
   
   With implementation of these recommendations, we believe the true character and form of the neutrino matrix can be illuminated, and its role  in the universe disclosed.



\end{twocolumn}
\begin{onecolumn}
\clearpage
\renewcommand{\theequation}{A-\arabic{equation}}
  \setcounter{equation}{0}  

In this Appendix, only the Executive Summaries of 
the Working Groups are presented.  The full text can be 
found at any of the four APS Divisional web sites.

\subsection{Executive Summary of Solar and Atmospheric 
Experiments Working Group}
{\bf participants:  }{\it 
H.~Back, J.N.~Bahcall, J.~Bernabeu, M.G.~Boulay, T.~Bowles, 
F.~Calaprice, A.~Champagne,  M.~Gai, C.~Galbiati, H.~Gallagher, 
C.~Gonzalez-Garcia, R.L.~Hahn, K.M.~Heeger, A.~Hime, C.K.~Jung, J.R.~Klein, 
M.~Koike, R.~Lanou, J.G.~Learned, K.~T.~Lesko, J.~Losecco, 
M.~Maltoni, A.~Mann, 
D.~McKinsey, S.~Palomares-Ruiz, C.Pe\~{n}a-Garay, 
S.T.~Petcov, A.~Piepke, 
M.~Pitt, R.~Raghavan, R.G.H.~Robertson, K.~Scholberg, 
H.W.~Sobel, T.~Takeuchi, R.~Vogelaar, L.~Wolfenstein  }
\medskip
\subsubsection{Introduction}
\label{sec:intro}
	Both the first evidence and the first discoveries of neutrino flavor
transformation have come from experiments which use neutrino beams provided
by Nature.  These discoveries were remarkable not only because they were
unexpected---they were discoveries in the purest sense---but that they were
made initially by experiments designed to do different physics.  Ray Davis's
solar neutrino experiment was created to study solar astrophysics, not the
particle physics of neutrinos.  The IMB, Kamiokande, and Super-Kamiokande experiments 
were hoping to
observe proton decay, rather than study the (ostensibly 
relatively
uninteresting) atmospheric neutrino flux.  That these experiments 
and their
successors have had such a great impact upon our view of 
neutrinos and the
Standard Model underscores two of the most important 
motivations for
continuing current and creating future solar and 
atmospheric neutrino
experiments: they are naturally sensitive to a 
broad range of physics (beyond
even neutrino physics), and they therefore have a great potential for the
discovery of what is truly new and unexpected.
	The fact that solar and atmospheric neutrino experiments use
naturally created neutrino beams raises the third important motivation---the
beams themselves are intrinsically interesting.  Studying atmospheric
neutrinos can tell us about the primary cosmic ray flux, and at high energies
it may bring us information about astrophysical sources of neutrinos (see
Report of Astrophysics Working Group) or perhaps even something about
particle interactions in regimes still inaccessible to accelerators.  For
solar neutrinos, the interest of the beam is even greater: as the only
particles which can travel undisturbed from the solar core to us, neutrinos
tell us details about the inner workings of the Sun.  The recent striking
confirmation of the predictions of the Standard Solar Model (SSM) are
virtually the tip of the iceberg:  we have not yet examined in an exclusive
way more than 99\% of the solar neutrino flux.  The discovery and
understanding of neutrino flavor transformation now allows us to return to
the original solar neutrino project---using neutrinos to understand the Sun.

	The fourth and perhaps strongest motivation for solar and atmospheric
neutrino experiments is that they have a vital role yet to play in exploring
the new physics of neutrinos.  The beams used in these experiments give
them unique sensitivity to some of the most interesting new phenomena.
The solar beam is energetically broadband, free of flavor backgrounds,
and passes through quantities of matter obviously unavailable to terrestrial
experiments.  The atmospheric beam is also broadband, but unlike the solar
beam it has the additional advantage of a baseline which varies from tens
of kilometers to many thousands.  

	The Solar and Atmospheric Neutrino Experiments Working Group  has 
chosen to 
focus on the following primary physics questions:
\begin{itemize}
\item {\it Is our model of neutrino mixing and oscillation complete, 
or are there other mechanisms at work?}
	To test the oscillation model, we must search for sub-dominant
effects such as non-standard interactions, make precision comparisons
to the measurements of other experiments in different regimes, and
verify the predictions of both the matter effect and vacuum oscillation.
The breadth of the energy spectrum, the extremely long baselines, and the
matter densities traversed by solar and atmospheric neutrinos make them
very different than terrestrial experiments, and hence measurements in all
three mixing sectors---including limits on $\theta_{13}$---can be compared
to terrestrial measurements and thus potentially uncover new physics.
\item {\it Is nuclear fusion the only source of the Sun's energy?}
	Comparison of the total energy output of the Sun measured in
neutrinos must agree with the total measured in photons, if nuclear
fusion is the only energy generation mechanism at work.  

\item {\it What is the correct hierarchical ordering of the neutrino masses?}

	Atmospheric neutrinos which pass through the Earth's core and mantle
will have their transformation altered due to the matter effect, dependent
upon the sign of the $\Delta m^2_{32}$ mass difference.  Future large scale
water Cerenkov experiments may be able to observe this difference in the ratio
of $\mu$-like to $e$-like neutrino interactions, while magnetized atmospheric
neutrino experiments may be able to see the effect simply by comparing the
number of detected $\nu_{\mu}$ to $\bar{\nu_{\mu}}$ events.
\end{itemize}

\subsubsection{Recommendations}

The highest priority of the Solar and Atmospheric Neutrino Experiment
Working Group is the development of a real-time, precision experiment
that measures the $pp$ solar neutrino flux.  A measurement of the $pp$
solar neutrino flux, in comparison with the existing precision measurements
of the high energy $^8$B neutrino flux, will demonstrate the transition
between vacuum and matter-dominated oscillations, thereby quantitatively
testing a fundamental prediction of the standard scenario of neutrino flavor
transformation. The initial solar neutrino beam is pure $\nu_e$, which
also permits sensitive tests for sterile neutrinos.  The $pp$ experiment
will also permit a significantly improved determination of $\theta_{12}$
and, together with other solar neutrino measurements, either a measurement
of $\theta_{13}$ or a constraint a factor of two lower than existing bounds.

	In combination with the essential pre-requisite experiments that
will measure the $^7$Be solar neutrino flux with a precision of 5\%, a
measurement of the $pp$ solar neutrino flux will constitute a sensitive test
for non-standard energy generation mechanisms within the Sun.  The Standard
Solar Model predicts that the $pp$ and $^7$Be neutrinos together constitute
more than 98\% of the solar neutrino flux. The comparison of the solar
luminosity measured via neutrinos to that measured via photons will test
for any unknown energy generation mechanisms within the nearest star.
A precise measurement of the $pp$ neutrino flux (predicted to be 92\% of
the total flux) will also test stringently the theory of stellar evolution
since the Standard Solar Model predicts the $pp$ flux with a theoretical
uncertainty of 1\%.
	We also find that an atmospheric neutrino experiment capable of
resolving the mass hierarchy is a high priority.  Atmospheric neutrino
experiments may be the only alternative to very long baseline accelerator
experiments as a way of resolving this fundamental question. Such an
experiment could be a very large scale water Cerenkov detector, or
a magnetized detector with flavor and antiflavor sensitivity.
	Additional priorities are nuclear physics measurements
which will reduce the uncertainties in the predictions of the Standard Solar
Model, and similar supporting measurements for atmospheric neutrinos (cosmic
ray fluxes, magnetic fields, etc.).  We note as well that the detectors
for both solar and atmospheric neutrino measurements can serve as
multipurpose detectors, with capabilities of discovering dark matter,
relic supernova neutrinos, proton decay, or as targets for long baseline 
accelerator neutrino experiments.

\subsection{Executive Summary of the Reactor Working Group}
{\bf participants:  }{\it E.~Abouzaid, K.~Anderson, G.~Barenboim, B.~Berger,
E.~Blucher, T.~Bolton, S.~Choubey,
J.~Conrad, J.~Formaggio, D.~Finley, 
P.~Fisher, B.~Fujikawa, M.~Gai,
M.~Goodman, A.~de Gouvea, N.~Hadley, R.~Hahn, G.~Horton-Smith,
R.~Kadel,  K.~Heeger,
J.~Klein, J.~Learned, M.~Lindner, J.~Link, K.-B.~Luk, R.~McKeown, I.
Mocioiu,
R.~Mohapatra, D.~Naples, J.~Peng, S.~Petcov,  J.~Pilcher, P.~Rapidis,
D.~Reyna, M.~Shaevitz, R.~Shrock, N.~Stanton, R.~Stefanski, R.~Yamamoto,
M.~Worcester }
\medskip
\subsubsection {Introduction}
The worldwide program to understand neutrino oscillations and
determine the mixing parameters, CP violating effects, and mass hierarchy
will require a broad combination of measurements.
Our group believes that a key element of this future neutrino program is
a multi-detector neutrino experiment (with baselines of $\sim 200$ 
m and $\sim 
1.5$ km)
with a sensitivity of $\sin^2 2 \theta_{13}=0.01$. 
In addition to oscillation physics,
the reactor experiment may provide interesting measurements
of $\sin^2 \theta_W$ at $Q^2=0$, neutrino couplings,
magnetic moments, and mixing with  sterile neutrino states.

$\theta_{13}$ is one of the twenty six parameters of the standard model,
the best model of electroweak interactions for energies below 100 GeV and,
as such, is worthy of a precision measurement independent of other
considerations.  A reactor experiment of the proposed sensitivity will
allow a measurement of $\theta_{13}$ 
with no ambiguities and significantly better
precision than any other proposed experiment, or will set limits indicating
the scale of future experiments required to make progress.
Figure~\ref{cpvs13} shows a comparison of the sensitivity of reactor
experiments of different scales with accelerator experiments for setting
limits on $\sin^2 2\theta_{13}$ if the mixing angle is very small, or for
making a measurement of $\sin^2 2\theta_{13}$ if the angle is observable.
A reactor experiment with a $1\%$ precision may also resolve the degeneracy
in the $\theta_{23}$ parameter when combined with long-baseline
accelerator experiments (see Fig.~\ref{cpvs13}).

\begin{figure}[htbp]
\begin{center}
\includegraphics[width=3.2in]{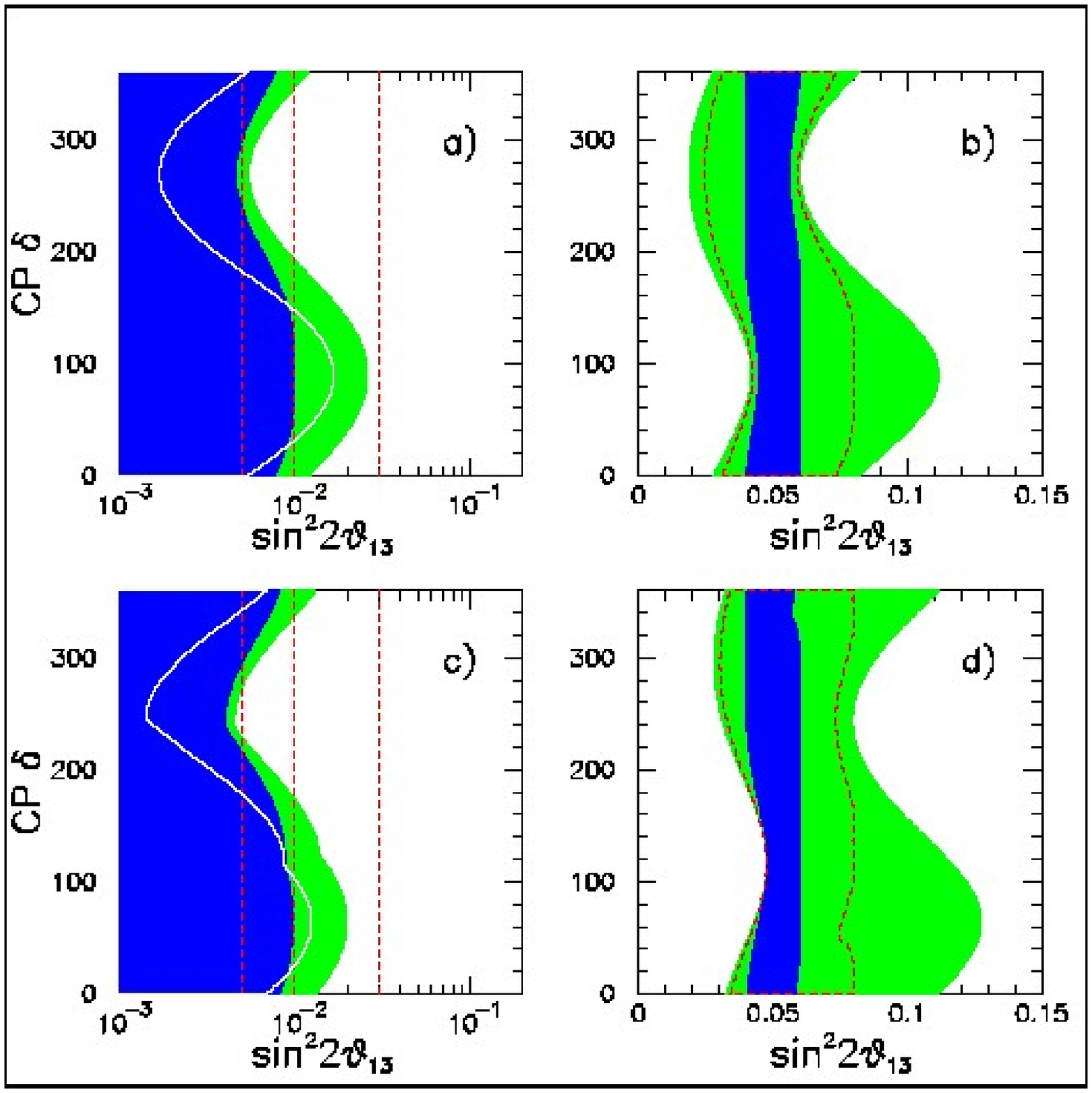}
\includegraphics[width=3.2in]{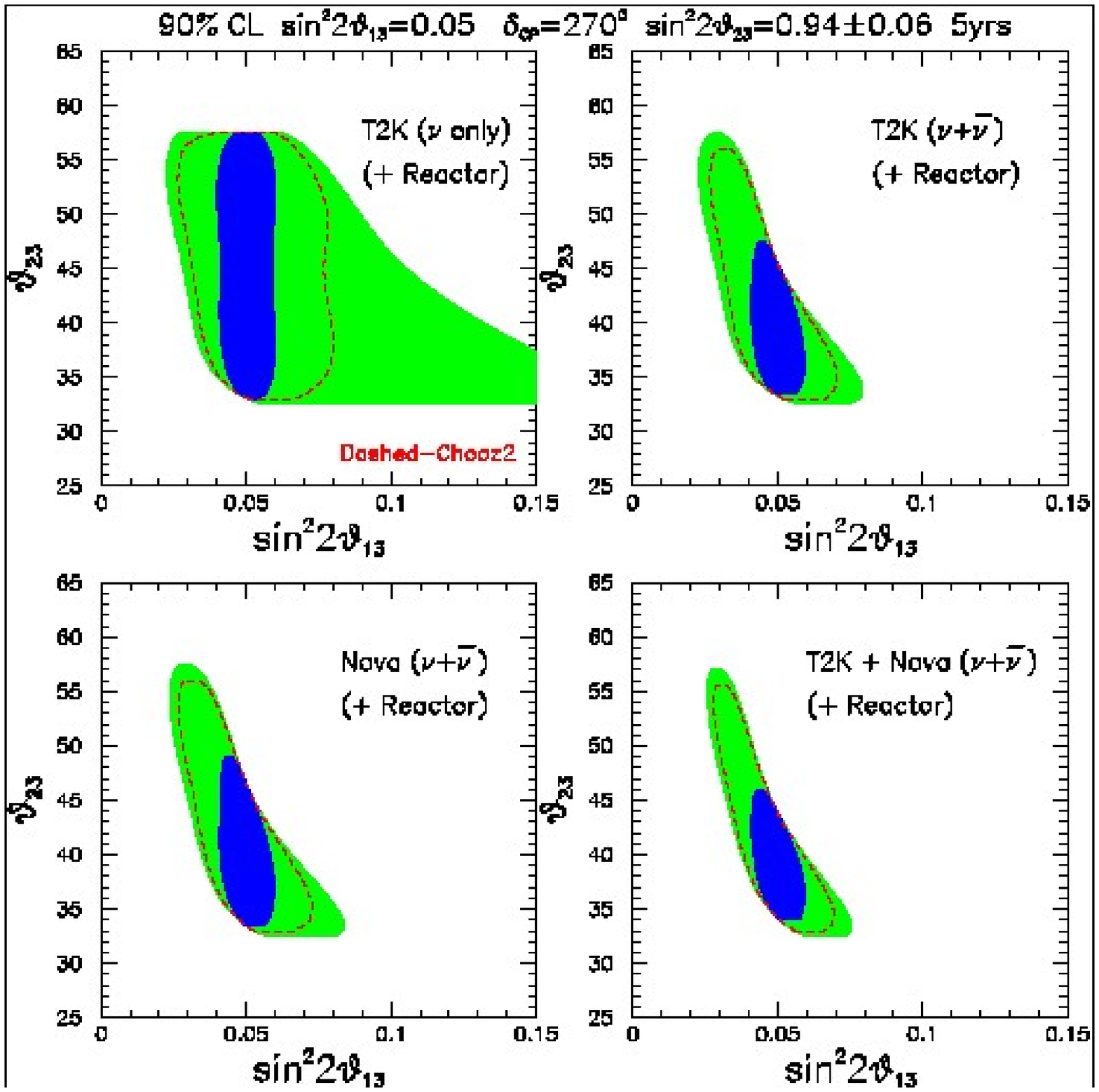}
\end{center}
\vspace*{-.2truein}
\caption{{\bf Left 4 Panels:} 90\% C.L. regions and 
upper limits for various oscillation
measurements for (a,c) $\sin^{2}%
2\theta_{13}=0$ and (b,d) 
$\sin^2 2 \theta_{13}=0.05$. 
The top (bottom) plots are for the T2K (Nova)
long-baseline experiments. The three vertical dashed lines in 
(a) and (c)
correspond to the
90\% C.L. upper limits of $0.005$, 0.01, and 0.03 possible with different
scales of reactor experiments. The green region (white curve) is the 90\%
C.L. allowed region for the two long-baseline experiments for a five year
neutrino-only run with nominal ($\times5$) beam rate, 
and the blue region gives
the combination of the five year long-baseline measurement with a reactor
experiment with sensitivity of $\sin^2 2\theta_{13}=0.01$; 
in (b) and (d), the dashed curves show how 
the combined measurement would be
degraded with a reactor experiment with sensitivity of 
$\sin^2 2\theta_{13}=0.03$. {\bf Right 4 Panels:} 90\% C.L. allowed 
regions for simulated data with 
oscillation parameters of $\sin^{2}2\theta_{13}=0.05$, $\theta_{23}=38^{o}$,
$\Delta m^{2}=2.5\times10^{-3}$ eV$^{2}$ and $\delta_{CP}=270^{o}$. The 
analysis
includes the restriction that $\sin^{2}2\theta_{23}=0.94\pm0.06$. 
The green
regions are for various combinations of 
the T2K and/or Nova
experiments for five years of running periods. The blue regions are the 90\%
C.L. allowed regions for the combination  of a reactor experiment with 
experiment. The dashed red lines show how the combined measurement would be
degraded with a reactor experiment with 3 times worse sensitivity.
}%
\label{cpvs13}%
\end{figure}

In combination with long-baseline measurements, a reactor experiment may
give early indications of CP violation and the mass hierarchy.
The combination
of  the T2K and Nova long-baseline experiments will be able to make
significant measurements of these effects if $\sin^2 2\theta_{13}>0.05$ 
and with
enhanced beam rates can improve their reach to the $\sin^2
2 \theta_{13}>0.02$ level.
If $\theta_{13}$ turns out to be smaller than 
these values, one will need other
strategies for getting to the physics.  Thus, an unambiguous reactor
measurement of $\theta_{13}$ 
is an important ingredient in planning the strategy
for the future neutrino program.

\subsubsection{Recommendations}
Our group has one highest priority recommendation:
\begin{itemize}
\item We recommend the rapid construction of a multi-detector
reactor experiment with a sensitivity of 0.01 for $\sin^2 2 \theta_{13}$.
\end{itemize}
Our other recommendations are the following:
\begin{itemize}
\item To help accomplish our highest priority recommendation, we recommend 
R\&D support necessary to prepare a full proposal.
\item We recommend continued support for the KAMLAND experiment. 
KAMLAND has made the best determination of $\Delta m_{12}^2$ to date, 
and will provide the best measurement for the foreseeable future. 
As the deepest running reactor experiment, it also provides 
critical information about cosmic-ray related backgrounds for future 
experiments.
\item We recommend the exploration of potential sites for a 
next-generation experiment at a distance of 70 km from an 
isolated reactor complex to make high precision measurements of $\theta_{12}$
and $\Delta m_{12}^2$.
\item We recommend support for development of future large-scale reactor 
$\theta_{13}$ experiments that fully exploit energy spectrum information.
\end{itemize}
\subsection{Executive Summary of the Superbeams Working Group}
{\bf participants:  }{\it 
C. Albright, D. Ayres, A. Bazarko, F. Bertrand,
G. Bock, D. Boehnlein, S. Brice, B. Brown,
L. Buckley-Geer, M. Campanelli
% ...
}

\subsubsection {Introduction}
As we seek the answers to the central questions in neutrino physics,
accelerator-based experiments will be
crucial for providing the necessary precision and sensitivity. There are
several physics questions which accelerator superbeam experiments will
address:
\begin {itemize}
\item {\it What is the mixing pattern among the neutrinos? Do the mixings
suggest some new fundamental mechanism which causes them to have unusual
values?}
\item {\it What is the mass hierarchy for the three known neutrinos?}
\item {\it Do neutrinos violate the symmetry CP?}
\item {\it Are there additional light neutrinos and do they participate
in oscillations with the three known neutrinos?}
\item {\it Do we understand the basic mechanism of neutrino oscillations?}

\item {\it Do neutrinos have measurable magnetic moments or other exotic
properties?}
\end {itemize}
Shorter-term
experiments will depend on existing accelerator capabilities. However, in 
the longer term it is now clear that we will require new or upgraded proton
accelerators capable of providing greater than a mega-Watt of proton power
for a neutrino superbeam. With such a driver, a rich new program of 
neutrino oscillation and other physics measurements will be possible.
\subsubsection {Recommendations}

{\bf I. Highest Priority Recommendation:}

\begin {itemize}
\item {\bf Build a new MW+ class proton driver, neutrino 
superbeam and very massive detector in the United States.} 
\end {itemize}
These are the necessary components for a 
complete set of precision measurements
on the oscillation parameters of interest. The key feature of these 
experiments is that they will provide ~1\% measurement of 
$\sin^2 2\theta_{23}$ and $\Delta m^2_{23}$ and sensitivity to 
$\sin^2 2\theta_{13}$ below 0.01 (depends on the other parameters). 
Should $\sin^2 2\theta_{13}$ be greater than about 0.01 these experiments
will also provide discovery and 
measurement capability for CP violation and, because of  the
long baselines, unique measurement capability of the mass hierarchy.
A very large multi-purpose detector located at an underground site will
permit not just long-baseline oscillation measurements but also 
measurements on solar and atmospheric neutrinos, a search for supernova
neutrinos and a search for proton decay. The new proton driver will enable
both long and short baseline oscillation experiments as well as a variety
of other neutrino experiments. It will also permit new precise muon and
hadron experiments as well as act as the essential first stage of a possible
future neutrino factory.

 {\bf II. Short-term Recommendations:}
\begin {itemize}
\item {\bf Significant design studies for a new proton driver facility have 
been completed over the last few years. We urge a rapid decision on this 
facility.} 
We expect that it will take roughly 8 years from now before
a new proton driver could be completed, if the decision to proceed and 
selection of the site is done soon. Moving now to decide on this machine
will permit the U.S. to have the leading program of neutrino measurements
in the following decade.
\item {\bf Increase proton intensity at Fermilab, roughly 
by about a factor of 2 in both 
the Booster and Main Injector neutrino beamlines
over the next few years.} 
Both the MINOS and Mini-BooNE experiments offer
exciting discovery and measurement potential in the next few years but their
capabilities depend critically on proton intensity. Roughly, we encourage
investment with a goal of delivering about $4\times10^{20}$ protons per
year at both 8 GeV and 120 GeV. 
\item {\bf We recommend the LSND result be tested with both neutrinos and 
anti-neutrinos.}
Mini-BooNE is currently using neutrinos to test the LSND result (which is 
$\overline{\nu_e}$ appearance in an initial beam of $\overline{\nu_\mu}$).
It is essential that this test be conclusive. Should Mini-BooNE not
confirm LSND with neutrinos, testing the result with anti-neutrinos
will be important. Improvements in proton intensity as discussed in the
preceding recommendation would permit Mini-BooNE to also test LSND
with anti-neutrinos.
\item {\bf We endorse the physics goals of a long-baseline $\nu_e$
appearance experiment using the existing NuMI beamline. 
We recommend development of this experimental 
program. A reactor neutrino experiment running in parallel will be 
complementary.}
Such an experiment should be roughly 10 times more sensitive than MINOS to 
$\nu_e$ appearance and being done in a timely manner would capitalize
on the considerable investment in NuMI. With a suitable detector, a
properly optimized appearance experiment could have good sensitivity
to $\theta_{13}$ and provide a unique relatively short-term opportunity to
determine the neutrino mass hierarchy via matter effects. That determination
would have important implications for fundamental neutrino properties as 
well as the requirements for future neutrinoless double beta decay 
experiments.
\end {itemize}
{\bf III. Long Term Strategy and Priorities:}
\begin {itemize}
\item {\bf Pursue a long-baseline neutrino program. 
The U.S. should focus on longer baseline experiments than are being 
considered in Japan or Europe (at present at least). The overall U.S. 
program (domestic and participation in experiments abroad) should form a 
coherent part of an international effort.}
Neutrino Superbeam experiments being planned in Japan and Europe have
baselines sufficiently short so that it is difficult to measure the matter
effects which can identify the mass hierarchy. This is a unique measurement
capability which we believe the U.S. experiment(s) should offer. In addition,
the U.S. experiments have the potential for providing the best sensitivity
to the oscillation parameters, including first measurement of $\nu_e$
appearance and discovery and measurement of CP violation in neutrino 
oscillations.
\item {\bf A massive detector will be necessary for the future long-baseline 
experiments. We recommend a study of the possible eventual connection between 
a neutrino superbeam with a massive multi-purpose detector.}
One can probably build the very large detector needed just for long baseline
experiments alone on the surface. However, the capabilities which such a
detector must have can permit a broad range of physics 
measurement capabilities
if located underground. We think it is essential to study the technology and
possible connections between the superbeam and multi-purpose underground
detector.
\item {\bf If LSND is confirmed, a whole new range of experiments should 
follow with possible programs at a variety of laboratories.}
If the LSND observation is correct, then there are light sterile neutrinos
which also participate in oscillations, or something even stranger yet.
This modifies the model of neutrino mixings in a way that requires us to
provide measurements to both establish the very nature of the mixing as
well as specific values of parameters. Long baseline experiments with the
capabilities we describe here will still be essential, but the
interpretation of their results may be different. In addition, new short
(or possibly medium) baseline experiments will be essential to study the
new physics phenomena in detail and build a new picture of neutrino physics.

\item {\bf Searches for exotic neutrino properties should be pursued with 
new 
superbeam experiments.}
Due to their special properties, neutrinos can be particularly sensitive
to a range of possible new physics from extra dimensions to violation
of equivalence principle to new very weak interactions. Relatively small
new short-baseline experiments are able to extend sensitivity to 
possible exotic physics associated with neutrinos and such experiments will
become better as higher intensity neutrino beams are available. A good
example of such a measurement is to search for an anomalously large
neutrino magnetic moment induced  by effects of extra dimensions.
Experiments extending such sensitivity by a factor of 10--100 are foreseen.

\item {\bf New high-precision cross-section experiments 
should be undertaken.}
Detailed understanding of neutrino interaction cross sections is 
important for
future oscillation measurements. Such measurements can also provide 
interesting
insight to QCD effects and effects of nuclear matter. Current understanding
of cross-sections (total, differential and exclusive final states) in the
GeV range, so important to oscillation experiments, is only at the tens of
percent level. Although near detectors can help to cancel some of the 
uncertainty in cross sections, the better and more precise solution is to
actually measure the cross sections better than currently known once and
for all! We encourage that the experiments necessary for this be carried out.

\end {itemize}

\subsection{Executive Summary of the Neutrino Factory and Beta 
Beam Experiments and Development Working Group}

{\bf participants:  }\textit{C.~Albright, 
V.~Barger, 
J.~Beacom, 
J.S.~Berg, 
E.~Black, 
A.~Blondel, S.~Brice, S.~Caspi, W.~Chou, M.~Cummings,
R.~Fernow,
D.~Finley, J.~Gallardo, S.~Geer, 
J.J.~Gomez-Cadenas,
M.~Goodman,
D.~Harris,
P.~Huber,
A.~Jansson,
C.~Johnstone, S.~Kahn, 
D.~Kaplan, 
H.~Kirk, T.~Kobilarcik, 
M.~Lindner, K.~McDonald, 
O.~Mena, 
D.~Neuffer, 
V.~Palladino, 
R.~Palmer, 
K.~Paul, 
P.~Rapidis, 
N.~Solomey, 
P.~Spampinato, 
Y.~Torun, 
K.~Whisnant, W.~Winter, M.~Zisman
\begin{center} 
and\\
 The Neutrino Factory and Muon Collider Collaboration
\end{center}
}
\subsubsection {Introduction}
\medskip
Two new types of facility have been proposed that could have a tremendous 
impact on future neutrino experiments---the Neutrino Factory and the 
Beta Beam facility. In contrast to conventional muon-neutrino beams, 
Neutrino Factory and Beta Beam facilities would provide a 
source of electron-neutrinos $(\nu_e)$ and -antineutrinos $(\bar{\nu}_e)$, 
with very low systematic uncertainties on the associated 
beam fluxes and spectra. The experimental signature for 
$\nu_e \to \nu_\mu$ transitions is extremely clean, with 
very low background rates. Hence, Neutrino Factories and Beta Beams 
would enable very sensitive oscillation measurements to be made. This 
is particularly true at a Neutrino Factory, which not only provides 
very intense beams at high energy, but also provides 
muon-neutrinos $(\nu_{\mu})$ 
and -antineutrinos $(\bar{\nu}_{\mu})$ in 
addition to electron-neutrinos $(\nu_e)$ and -antineutrinos 
$(\bar{\nu}_e)$. This would facilitate 
a large variety of complementary oscillation measurements in a 
single detector, and dramatically improve our ability to test the 
three-flavor mixing framework, measure CP violation in the 
lepton sector (and perhaps determine the neutrino mass hierarchy), 
and, if necessary, probe extremely 
small values of the mixing angle $\theta_{13}$. 
 
At this time, we do not know the value of $\theta_{13}$. If 
$\sin^{2}%
2\theta_{13}<0.01$, much of the basic neutrino oscillation physics program 
will be beyond the reach of conventional neutrino beams. In this case 
Neutrino Factories
and Beta Beams offer the only known way to pursue the desired physics 
program.
The sensitivity that could be achieved 
at a Beta Beam facility presently looks
very promising, but is still being explored. 
In particular, the optimum Beta
Beam energy is under discussion. Low energy Beta Beam measurements would
complement Superbeam measurements, but would achieve a $\theta_{13}$ 
sensitivity that does
not appear to be competitive with 
that of a Neutrino Factory. Higher energy
Beta Beams may approach the sensitivity possible with a Neutrino Factory,
although systematics issues need further study. Thus, while a Beta Beam
facility may have a significant role to play in the future global neutrino
program,  more work must be done on its design, development, cost estimate,
and physics sensitivity to validate its potential. 
We note that, due to very limited resources, there has been no
significant activity in the U.S. on Beta Beams. Progress on Beta Beam 
development being made in Europe should be followed, 
especially if the higher energy solution continues to look favorable.
An impressive Neutrino Factory R\&D effort has been ongoing in the U.S. and

elsewhere over the last few years, 
and significant progress has been made
toward optimizing the design, developing and testing the required accelerator
components, and significantly reducing the cost, even during the current
Study. (Although a full engineering study is required, we have preliminary
indications that the unloaded cost of a Neutrino Factory 
facility based on an
existing Superbeam proton driver and target station can be reduced
substantially compared with previous estimates.) Neutrino Factory R\&D has
reached a critical stage in which support is required for two key
international experiments (MICE and Targetry) and a third-generation
international design study. If this support is forthcoming, 
a Neutrino Factory
could be added to the Neutrino Physics roadmap in about a decade. 
Given the present uncertainty about the 
size of $\theta_{13},$ \textit{it is critical to 
support an ongoing and increased U.S. investment in 
Neutrino Factory accelerator R\&D to maintain this technical option}. 
A Neutrino
Factory cannot be built without continued and increased
support for its development. We note that the 2001 HEPAP Report 
advocated an annual U.S. investment of \$8M on Neutrino Factory R\&D. The 
present support is much less than this. Since R\&D on the design of 
frontier accelerator facilities takes many years, 
support must be provided \textit{now} to have an
impact in about a decade.

\subsubsection{Recommendations}
Accelerator R\&D is an essential 
part of the ongoing global neutrino program.
Limited beam intensity is already constraining the neutrino physics program,
and will continue to do so in the future. More intense and new types of
neutrino beams would have a big impact on the future neutrino program. A
Neutrino Factory would require a Superbeam-type MW-scale  proton source. We
thus encourage the rapid development of a Superbeam-type proton source. 
The Neutrino Factory and Beta Beam Working Group's 
specific recommendations are:
\begin{itemize}
\item \textbf{\textit{We recommend that the ongoing 
Neutrino Factory R\&D in the U.S.
be given continued encouragement and financial support.}} We note that the
HEPAP Report of 2001 recommended an annual support level of \$8M for Neutrino
Factory R\&D, and this level was considered minimal to keep the R\&D effort
viable.

In addition, and consistent with the above recommendation,
\begin{enumerate}
\item \textbf{\textit{We recommend that the U.S. funding agencies find 
a way to
support the international Muon Ionization Cooling Experiment (MICE), in
collaboration with European and Japanese partners.}} We note that MICE 
now has
scientific approval at the Rutherford Appleton Laboratory in 
the UK, and will
require significant U.S. participation. This has been identified as an
important experiment for the global Neutrino Factory R\&D program. A timely
indication of U.S. support for MICE is needed to 
move the experiment forward. 
\item \textbf{\textit{We recommend that support be found to ensure that the
international Targetry R\&D experiment proceeds 
as planned.}} We note that this
R\&D activity is crucial for the short-, medium-, and long-term neutrino
programs, and for other physics requiring high-intensity beams. 
\item \textbf{\textit{We recommend that a World Design Study, aimed at solidly
establishing the cost of a cost-effective Neutrino Factory, be supported at
the same level as Studies I and II.}} We note that the studies done here
suggest that the cost of a Neutrino Factory would be significantly less than
estimated for Studies I and II. This makes a Neutrino Factory a very
attractive ingredient in the global neutrino roadmap.
\end{enumerate} 
\item \textbf{\textit{We recommend that progress on Beta Beam development be
monitored, and that our U.S. colleagues cooperate fully with their EU
counterparts in assessing how U.S. facilities might play a role in such a
program.}} We note that there is no significant U.S. R\&D effort on Beta Beams
due to our limited R\&D resources. Insofar as an intermediate energy solution
is desirable, however, the Beta Beam idea is potentially of interest to the
U.S. physics community.
\end{itemize}
\subsection{Executive Summary of the
Neutrinoless Double Beta Decay and 
Direct Searches for Neutrino Mass Working Group}
\subsubsection {Introduction}
{\bf participants:  }{\it  
C.~Aalseth,
H.~Back,
L.~Dauwe,
D.~Dean,
G.~Drexlin,
Y.~Efremenko,
H.~Ejiri,
S.~Elliott,
J.~Engel,
B.~Fujikawa,
R.~Henning,
G.W.~Hoffmann,
K.~Lang,
K.~Lesko,
T.~Kishimoto,
H.~Miley,
R.~Norman,
S.~Pascoli,
S.~Petcov,
A.~Piepke,
W.~Rodejohann,
D.~Saltzberg,
S.~Sutton,
P.~Vogel,
R.~Warner,
J.~Wilkerson,
and L.~Wolfenstein.

}
\medskip
The physics addressed by this research program seeks to answer 
many of the Study's questions:  
\begin{enumerate}
\item Are neutrinos their own anti-particles?
\item What are the masses of the neutrinos?
\item Do neutrinos violate the symmetry CP?
\item Are neutrinos the key to the understanding of 
the matter-antimatter asymmetry of the 
Universe?
\item What do neutrinos have to tell 
us about the intriguing proposals for new models of physics?
\end{enumerate}
Only the research covered within this working group can 
answer the first  and second of these fundamental questions.
Among the ways to measure the neutrino mass, 
three are notable because they are
especially sensitive: double-beta decay, tritium beta decay, 
and cosmology. Consequently, we have focused our report 
and recommendations on them. 
$\bullet$ Observation of the  neutrinoless double-beta 
decay ($0\nu\beta\beta$) 
would prove that the total lepton number is not conserved
and would establish a non-vanishing
neutrino mass of Majorana nature. 
In other words, observation of the $0\nu\beta\beta$ decay, independent
of its rate, would show that neutrinos, unlike all the other
constituents of matter, are their own antiparticles.
There is no other realistic way
to determine the nature --- Dirac or Majorana --- of massive neutrinos. 
This would be a discovery of major importance, with impact not only on
this fundamental question, but also on the determination of the absolute
neutrino mass scale, on the pattern of neutrino masses, and possibly 
on the problem of CP violation in the lepton sector.
There is  consensus on this basic point,
which we translate into the recommendations on
how to proceed with experiments dedicated to the
search for  $0\nu\beta\beta$ decay, and on how to fund them.
To reach our conclusion, we  have to consider past achievements,
the size of previous experiments, and the existing proposals. 
There is a considerable community
of physicists worldwide as well as in the US interested in pursuing the
search for the  $0\nu\beta\beta$ decay. Past experiments were
of relatively modest size. Clearly, the scope
of future experiments should be considerably larger, and will require
advances in experimental techniques, larger collaborations 
and additional funding.
In terms of $\langle m_{\beta\beta} \rangle$, the effective neutrino Majorana
mass that can be extracted 
from the observed  $0\nu\beta\beta$ decay rate, there are three ranges
of increasing sensitivity, related to known neutrino-mass scales 
of neutrino oscillations.
$\bullet$ The $\sim$100-500 meV  $\langle m_{\beta\beta} \rangle$
range corresponds to the quasi-degenerate
spectrum of neutrino masses. The motivation for reaching 
this scale has been strengthened by the recent
claim of an observation of $0\nu\beta\beta$ decay in $^{76}$Ge;
a claim that obviously requires further investigation. To reach this scale
and perform reliable measurements,
the size of the experiment should be approximately 
200 kg of the decaying isotope, 
with a corresponding reduction of the background.
This quasi-degenerate scale is achievable in the relatively
near term, $\sim$ 3-5 years.
Several groups with considerable US participation have well established
plans to build $\sim$ 200-kg devices that could scale straight-forwardly
to 1 ton (Majorana using $^{76}$Ge, Cuore using  $^{130}$Te, 
and EXO using  $^{136}$Xe). There are also 
other proposed experiments worldwide which
offer to study a number of other isotopes and
could reach similar sensitivity after further R\&D.
Several among them ({\it e.g.} Super-NEMO, MOON) have US participation.

By making measurements in several
nuclei the uncertainty arising from the nuclear
matrix elements would be reduced. The development of different
detection techniques, and measurements in several nuclei, 
is invaluable for establishing the existence
(or lack thereof) of the  $0\nu\beta\beta$ decay at this
effective neutrino mass range.       
$\bullet$ The $\sim$20-55 meV range arises from the atmospheric
neutrino oscillation results.  Observation of 
$\langle m_{\beta\beta} \rangle$ at this mass scale 
would imply the inverted
neutrino mass hierarchy or the normal-hierarchy 
$\nu$ mass spectrum very near the quasi-degenerate
region.
If either this or the quasi-degenerate spectrum is established,
it would be invaluable not only for the understanding of the 
origin of neutrino mass, but also as input 
to the overall neutrino physics program 
(long baseline oscillations, search for
CP violations, search for neutrino mass in tritium beta decay
and astrophysics/cosmology, etc.)  
To study the 20-50 meV mass range will require about 1
ton of the isotope mass, a  challenge of its own.
Given the importance, and the points discussed above, more 
than one experiment of that size is desirable.
$\bullet$ The $\sim$2-5 meV range arises from the solar
neutrino oscillation results and will almost certainly lead
to the  $0\nu\beta\beta$ decay, provided neutrinos are
Majorana particles. 
To reach this goal will require $\sim$100 tons of the 
decaying isotope,
and no current technique provides such a leap in sensitivity. 
The qualitative physics results that arise from an observation
of $0\nu\beta\beta$ decay are profound. Hence, the program described above
is vital and fundamentally important
even if the resulting $\langle m_{\beta\beta} \rangle$ would be 
rather uncertain in value.
However,  by making measurements in several
nuclei the uncertainty arising from the nuclear
matrix elements would be reduced.

Unlike double-beta decay, beta-decay endpoint measurements  search
for a kinematic effect due to neutrino mass and  thus are 
``direct searches'' for neutrino mass. This technique,
which is essentially free of theoretical assumptions
about neutrino properties, is not just complementary. In fact,
both types of measurements will be required to fully untangle 
the nature of the neutrino mass.
Excitingly, a very large new beta spectrometer is being built in Germany. 
This KATRIN experiment
has a design sensitivity approaching 200 meV. 
If the neutrino masses are quasi-degenerate, as would
be the case if the recent double-beta decay claim proves true, 
KATRIN will see the effect. In 
this case the $0\nu\beta\beta$-decay experiments 
can provide, in principle, unique information
about CP-violation in the lepton sector, associated with Majorana 
neutrinos.

Cosmology can also provide crucial information on the sum of the
neutrino masses. This topic is summarized in 
a different section of the report, but it 
should be mentioned here that the next generation of measurements 
hope to be able to observe a sum of neutrino masses as small as 40 meV. 
We would like to emphasize the complementarity 
of the three approaches, $0\nu\beta\beta$ ,
$\beta$ decay,
and cosmology. 

\subsubsection{Recommendations}
We conclude that such a double-beta-decay program can 
be summarized as having three
components and our recommendations
can be summarized as follows:
\begin{enumerate}
\item A substantial number (preferably more
than two) of 200-kg scale experiments (providing the capability to
make a precision measurement
at the quasi-degenerate mass scale) with large US participation
should be supported as soon as possible. \\
\item Concurrently, the development
toward $\sim$1-ton experiments ({\it i.e.} sensitive to $\sqrt{\Delta
m_{\rm atm}^{2}}$)
should be supported, primarily as expansions of the 200-kg experiments.
The corresponding plans for the procurement
of the enriched isotopes, as well as for the development of a suitable
underground facility, should be carried out. The US funding agencies
should set up in a timely manner a mechanism to review and compare
the various proposals for such experiments which span
research supported by the High Energy and Nuclear Physics offices of DOE
as well as by NSF. \\
\item A diverse R\&D program developing
additional techniques should be supported.\\
\end{enumerate}
$\bullet$ In addition to double-beta decay, other techniques for
exploring the neutrino mass
need to be pursued also. We summarize these recommendations as follows.
\begin{enumerate}
\item Although KATRIN is predominately a European effort, there
is significant US participation.
The design and construction of this experiment is proceeding well and
the program should
continue to be strongly supported.
\item Research and development of other techniques for observing the
neutrino mass kinematically
should be encouraged.
\end{enumerate}
\subsection{Executive Summary of the Neutrino Astrophysics and 
Cosmology Working Group}
{\bf participants: }
{\it
B. Balantekin,
S. Barwick,
J. Beacom,
N. Bell,
G. Bertone,
D. Boyd,
L. Chatterjee,
M.-C. Chen,
V. Cianciolo,
S. Dodelson,
M. Dragowsky,
J.L. Feng,
G. Fuller,
E. Henley,
M. Kaplinghat,
A. Karle,
T. Kattori,
P. Langacker,
J. Learned,
J. LoSecco,
C. Lunardini,
D. McKay,
M. Medvedev,
P. M\'esz\'aros,
A. Mezzacappa,
I. Mocioiu,
H. Murayama,
P. Nienaber,
K. Olive,
S. Palomares-Ruiz,
S. Pascoli,
R. Plunkett,
G. Raffelt,
T. Stanev,
T. Takeuchi,
J. Thaler,
M. Vagins,
T. Walker,
N. Weiner,
B.-L. Young}

\medskip
\subsubsection {Introduction}
In 2002, Ray Davis and Masatoshi Koshiba were awarded the Nobel Prize
in Physics ``for pioneering contributions to astrophysics, in
particular for the detection of cosmic neutrinos.''  However, while
astronomy has undergone a revolution in understanding by synthesizing
data taken at many wavelengths, the universe has only barely been
glimpsed in neutrinos, just the Sun and the nearby SN 1987A.  An
entire universe awaits, and since neutrinos can probe astrophysical
objects at densities, energies, and distances that are otherwise
inaccessible, the results are expected to be particularly exciting.
Similarly, the revolution in quantitative cosmology has heightened the
need for very precise tests that are possible only with neutrinos, and
prominent among them is the search for the effects of neutrino mass,
since neutrinos are a small but known component of the dark matter.
The Neutrino Astrophysics and Cosmology Working Group put special emphasis on
the following primary questions of the Neutrino Study; there are also
strong connections to the other questions as well.
\begin{itemize}
\item {\it What is the role of neutrinos in shaping the universe?}
\item {\it Are neutrinos the key to the understanding of the

matter-antimatter asymmetry of the universe?}
\item {\it What can neutrinos disclose about

    the deep interior of astrophysical 
objects,  and about the mysterious sources of very high energy cosmic

    rays?}
\end{itemize}
\subsubsection{Recommendations}
Our principal recommendations are:
\begin{itemize}
\item {\bf We strongly recommend the development of experimental

techniques that focus on the detection of astrophysical neutrinos,

especially in the energy range above $10^{15}$ eV.}
We estimate that the appropriate cost is less than \$10 million to
enhance radio-based technologies or develop new technologies for high
energy neutrino detection.  The technical goal of the next generation
detector should be to increase the sensitivity by factor of 10, which
may be adequate to measure the energy spectrum of the expected GZK
(Greisen-Zatsepin-Kuzmin) neutrinos, produced by the interactions of
ultra-high energy cosmic ray protons with the cosmic microwave
background (Fig.~\ref{AstroFig1}).  The research and development 
phase for these experiments
is likely to require 3-5 years.
\begin{figure}[!h]
\begin{center}
\includegraphics[,width=4in]{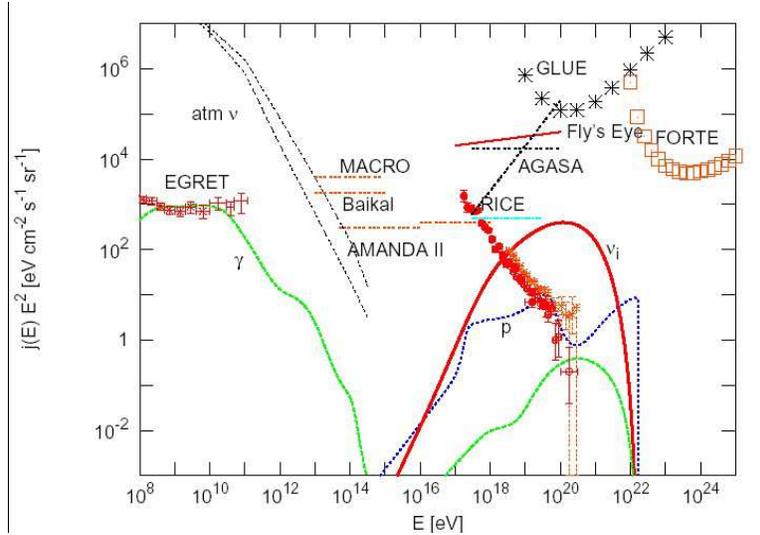}
\caption{Results are shown for the neutrino flux (solid red line)
predicted by a model of D.V. Semikoz and G. Sigl (JCAP 0404:003 (2004)
[hep-ph/0309328]), compared to existing limits (horizontal lines
labeled by the experiments).  This model is chosen to produce the
largest neutrino flux compatible with both the cosmic ray (red data
points, blue dotted lines) and gamma ray data (red data points, green
dashed lines), yet it remains beyond the reach of current experiments.
A new generation of experiments is needed to test these very
important predictions, as well as to begin to survey the ultra-high
energy universe for new sources.
\label{AstroFig1}}
\end{center}
\end{figure}
\item {\bf We recommend support for new precision measurements of
neutrino-nucleus cross sections in the energy range of a few tens of
MeV.}
We estimate that measurements of neutrino cross-section recommended by
this working group can be accomplished for less than \$10 million,
with R\&D requiring \$0.5 million for one year.  Construction will
require two additional years.
\item {\bf We recommend that adequate resources be provided to allow
existing large-volume solar, reactor, proton decay, and high energy
neutrino telescopes to observe neutrinos from the next supernova
explosion and participate in a worldwide monitoring
system. Furthermore, future large-volume detectors should consider the
detection of supernova neutrinos an important science goal and plan
accordingly.}
We anticipate that the investment to insure that large volume
detectors maintain sensitivity to galactic supernovae, as well as the
diffuse supernova neutrino background from all supernovae, will be
less than \$10 million over the next 5 years.  New large volume
detectors expected for long-baseline, reactor, proton-decay, solar,
and high energy neutrino detectors should consider new ideas to
enhance the capabilities for the detection of supernova neutrinos. The
cost is not possible to determine at this time.
\end{itemize}
Our principal endorsements are:
\begin{itemize}
\item {\bf We enthusiastically support continued investment in a
vigorous and multi-faceted effort to precisely (but indirectly)
measure the cosmological {\em neutrino} background through its effects
on big-bang nucleosynthesis, the cosmic microwave background, and the
large-scale structure of galaxies; in particular, weak gravitational
lensing techniques offer a very realistic and exciting possibility of
measuring neutrino masses down to the scale indicated by neutrino
oscillations.}
\item {\bf We enthusiastically support theoretical and computational
efforts that integrate the latest results in astronomy, astrophysics,
cosmology, particle physics, and nuclear physics to constrain the
properties of neutrinos and elucidate their role in the universe.}
\item {\bf We enthusiastically support the scientific goals of the
current program in galactic and extra-galactic neutrino astrophysics
experiments, including Super-Kamiokande, AMANDA, and NT-200 deployed
in Lake Baikal.  Furthermore, we endorse the timely completion of
projects under construction, such as IceCube, undersea programs in the
Mediterranean, ANITA, and AUGER.}
\item {\bf Though solar neutrinos were not in our purview, we endorse
the conclusion of the Solar/Atmospheric Working Group that it is
important to precisely measure solar neutrinos, and strongly support
the development of techniques which could also be used for direct
dark matter detection.}
\end{itemize}
\subsection{Executive Summary of the Theory Discussion Group}
{\bf participants:  }{\it   S. Antusch, K. S. Babu, G. Barenboim,
Mu-Chun Chen, S. Davidson, A. de Gouv\^ea, P. de Holanda, B.
Dutta, Y. Grossman, A. Joshipura, J. Kersten, Y. Y.
Keum, S. F. King, P. Langacker, M. Lindner, W. Loinaz, I.
Masina,  I. Mocioiu, S. Mohanty, R. N. Mohapatra, H. Murayama,
Silvia Pascoli, S. Petcov, A. Pilaftsis, P. Ramond, M.~Ratz,
W.~Rodejohann, R. Shrock, T. Takeuchi, T. Underwood, F. Vissani,
L. Wolfenstein }
\medskip
\subsubsection{Introduction}
Various oscillation experiments, from solar and
atmospheric to reactor and accelerator neutrinos have conclusively
established that neutrinos have mass and mix.
Thanks to these experiments, we now know: (i) the rough magnitude
of the leptonic mixing
angles (two of the three are large and a third one relatively
small) and
(ii) that the masses of all three neutrino species are exceedingly small
compared to charged fermion masses. This very small amount of
information has already served as  a  source of great excitement as 
it provides the
first (and currently only) evidence of physics beyond the standard model.
The discovery of neutrino masses also raises hope that one of the
fundamental mysteries
of the cosmos -- why there is more matter than anti-matter? -- may be
eventually resolved  through a better understanding of neutrinos.
There are, however, other fundamental neutrino properties,
related to their masses, about which we do not have information yet.
To elevate our knowledge of neutrinos to
the same level as that of the quarks,
the theory discussion group has attempted
to provide a prioritized list of the essential properties of neutrinos
needed for this purpose.
This would surely shed essential
light on the nature
of the new physics beyond the standard model as well as, perhaps,  the
origin of matter.
The key questions whose answers we do not know are:
  
\begin{enumerate}
\item  Are neutrinos their own anti-particles?
\item What is the pattern of neutrino masses ?
\item  Is there CP violation in the leptonic sector?
\item  Are there additional neutrino species as may be hinted by the LSND
experiment?
\end{enumerate}
On the theoretical side, while there are several different ways to
understand small neutrino masses, the seesaw mechanism, which introduces a
set of heavy ``right-handed neutrinos,'' appears to be the most appealing.
Existing data do not provide any way to verify if this idea is correct.
A key question here is whether the seesaw scale is near the grand
unification scale where all forces are expected to unify or much lower.

Before listing our
recommendations, we very briefly
discuss some of what we should learn from the results of various future
neutrino experiments:
{\bf (i) Searches for neutrinoless double beta decay:}
A positive signal would teach us that lepton number (or more precisely
the $B-L$ quantum number), which is an accidental symmetry of the standard
model in the absence of neutrino masses, is violated. This would provide
fundamental information, and would serve as a crucial
milestone in searches for new physics.
The popular seesaw mechanism predicts that neutrinos are
their own antiparticles, and the observation of
neutrinoless double beta decay would solidify it as the leading candidate
explanation for the origin of neutrino masses.
The observation of a positive signal in the foreseeable future would also
also imply the quasi-degenerate or inverted hierarchy for the neutrino
masses. The
quasi-degenerate pattern would suggest some special mechanism
  
for mass generation, possibly type II (Higgs triplet) seesaw, such
  
as can emerge in SO(10) grand unified theories (GUTs).
On the other hand, the absence of evidence for neutrinoless double beta

decay
would rule out the inverted and quasi-degenerate mass-hierarchies,
if the experiments reach an ultimate sensitivity of
$<m_{ee}>\simeq 15-50$ meV and if neutrinos are Majorana
particles. Furthermore,
  
if at the same time KATRIN observes a positive signal, we would learn that
neutrinos are Dirac fermions. This fact would have far reaching
implications for theory. It would, for example, contradict the predictions
of the seesaw theory.

{\bf (ii) Determination of  the mass hierarchy:}
This can obtained, for example,  from long baseline oscillation
experiments
An inverted mass hierarchy ($m_3^2 \ll m_1^2,m_2^2$), may be interpreted
to mean that leptons obey a new (only slightly broken)
symmetry: $L_e-L_\mu-L_\tau$, which would raise doubts about quark-lepton
symmetry, which is a fundamental ingredient of GUTs, such as SO(10).
A normal mass hierarchy ($m_3^2\gg m_1^2,m_2^2$), on the other hand,
is expected in generic seesaw models, including most SO(10) GUT
that address fermion masses and mixing.
{\bf (iii) Measurement of $\theta_{13}$:}
The next most important search item is the magnitude of $\theta_{13}$,
which can be obtained, for example,  from reactor neutrino experiments
as well as long baseline accelerator neutrino
experiments.
  
$\theta_{13}$ turns out to be one of the most clear discriminators among
various models
of neutrino masses. Simple symmetry arguments suggest that there
are two possible ranges for $\theta_{13}$: $\theta_{13}\simeq
\sqrt{\Delta m^2_{\odot}/\Delta m^2_{\rm atm}}\geq 0.1$ or
$\theta_{13}\simeq \Delta m^2_{\odot}/\Delta m^2_{\rm atm}\simeq0.04 $.
Of course, the magnitude of $\theta_{13}$ also determines whether
other fundamental questions (including ``is there leptonic CP
violation?" and ``what is the neutrino mass hierarchy?'') can be
experimentally addressed via neutrino oscillations.

{\bf (iv) CP violation and origin of matter:}

One may argue that CP violation in the leptonic
sector is expected, as strongly suggested by the presence of a large CP
phase in the quark sector. We believe, however, that detailed experimental
studies are required in order to determine the mechanism for leptonic
CP-violation (assuming it exists!).
The observation of leptonic CP-violation would enhance the possibility
that the matter asymmetry of the Universe was generated in
the lepton sector by demonstrating that CP violation exists
among leptons. However, there is no unambiguous connection:
the absence of CP-invariance violation in the light neutrino sector, for
example, would {\sl not} imply that
enough baryon asymmetry cannot be generated via the leptogenesis mechanism.
It turns out, however,  that models for leptogenesis generically imply
observable CP-invariance violation in the leptonic sector.
{\bf (v) Extra neutrinos:}
If the LSND anomaly is confirmed by MiniBooNE, a substantial change in
our understanding of high energy physics
will be required. One potential interpretation of the LSND anomaly is to
postulate the existence of (at least one) extra, ``sterile'' neutrino. This
would be a very concrete hint for new physics, beyond the traditional
seesaw, GUTs, etc.
If MiniBoone confirms the LSND anomaly, the most important task will be to
explore the nature of this phenomenon. It may turn out that LSND (and
MiniBooNE) have
uncovered some even more exotic phenomenon.
{\bf (vi) Other issues:}
Precision measurements of the solar neutrino spectrum can also provide
useful information about the detailed nature of matter effects on neutrino
propagation in the Sun as well as sources of energy generation there.
Similarly reactor searches for magnetic moment of neutrinos can also
provide signals of physics beyond the standard model such as
possible extra dimensions or new physics at TeV scale.

In this Working Group, approaches that focus on the following
primary physics questions are addressed:
\begin{itemize}
\item {\it Is our model of neutrino mixing and oscillation complete,
or are there other mechanisms at work?}

To test the oscillation model, we must search for sub-dominant
effects such as non-standard interactions, make precision comparisons
to the measurements of other experiments in different regimes, and
verify the predictions of both the matter effect and vacuum oscillation.
The breadth of the energy spectrum, the extremely long baselines, and the
matter densities traversed by solar and atmospheric neutrinos make them
very different than terrestrial experiments, and hence measurements in all
three mixing sectors---including limits on $\theta_{13}$---can be compared
to terrestrial measurements and thus potentially uncover new physics.
\item {\it Is nuclear fusion the only source of the Sun's energy, and is it
a steady state system?}

Comparison of the total energy output of the Sun measured in
neutrinos must agree with the total measured in photons, if nuclear
fusion is the only energy generation mechanism at work.  In addition,
the comparison of neutrino to photon luminosities will tell us whether
the Sun is in an approximately steady state by telling us whether the
rate of energy generation in the core is equal to that radiated through
the solar surface---the heat and light we see today at the solar surface
was created in the interior $\sim$ 40,000 years ago, while the neutrinos
are just over eight minutes old.

\item {\it What is the correct hierarchical ordering of the neutrino masses?}

Atmospheric neutrinos which pass through the Earth's core and mantle
will have their transformation altered due to the matter effect, dependent
upon the sign of the $\Delta m^2_{32}$ mass difference.  Future large scale
water Cerenkov experiments may be able 
to observe this difference in the ratio
of $\mu$-like to $e$-like neutrino interactions, while magnetized atmospheric
neutrino experiments may be able to see the effect simply by comparing the
number of detected $\nu_{\mu}$ to $\bar{\nu_{\mu}}$ events.
\end{itemize}
\subsubsection{Recommendations}
We very strongly recommend the following experiments, that will shed
light on the issues discussed above. We make the conservative assumption
that MiniBooNE will not confirm the LSND anomaly:
\begin{enumerate}
\item Double beta decay searches, which will shed light on whether
neutrinos are their own anti-particles;
\item Oscillation experiments capable of precisely measuring
all oscillation parameters, including the neutrino mass hierarchy,
$\theta_{13}$ and, ultimately, CP-violation;
\item Finally, we recommend that all resources be provided to Mini-Boone
until a satisfactory resolution of the LSND puzzle is obtained.
\end{enumerate}

\clearpage
\renewcommand{\theequation}{B-\arabic{equation}}
\renewcommand{\thesection}{B}
  \setcounter{equation}{0}  

\section{APS Study Origins, Committees, Glossary}

\subsection{The APS Multi-Divisional Neutrino Study}   

	To answer the very interesting questions raised by the discovery of neutrino mass, an effective, coherent strategy is needed. To foster the development of such a strategy, the American Physical Society's Divisions of Nuclear Physics and of Particles and Fields, together with the Divisions of Astrophysics and the Physics of Beams, have sponsored this yearlong Study on the Physics of Neutrinos. The study has endeavored to identify the most important open questions, to evaluate the physics reach of various proposed ways of answering them, and to determine an effective, fruitful U.S. role within a global experimental program. An important --  if challenging --  goal of the study has been to achieve  consensus regarding the future of neutrino physics. 

		A central element of the study has been its Working Groups, each defined by an experimental approach to answering the outstanding questions (see Table~\ref{tab:committees}). After the study's organizational meeting, held in December, 2003 at Argonne National Laboratory, the working groups carried out their activities autonomously, interacting with one another when appropriate to compare the different approaches to answering a given physics question, and to coordinate the attacks on related questions. The working groups presented their findings at the final joint meeting of the study, held in June, 2004 in Snowmass, Colorado. Those findings are now embodied in the Working Group Reports, the executive summaries of which appear in Appendix A of the present document.  The full texts may be found at {\bf http://www.interactions.org/neutrinostudy}. The meeting in Snowmass featured extensive discussion of the working group recommendations and of the study participants' opinions. 

	With the working group findings and the discussion in Snowmass as input, a Writing Committee (see Table~\ref{tab:committees}) has created the present final report of the study. This report, {\em The Neutrino Matrix}, is meant to integrate the working group findings into a coherent plan for the future that reflects the consensus that was evident in Snowmass. 

Overall guidance of the study has been provided by its Organizing Committee (see Table~\ref{tab:committees}). This committee planned the course of the study, and watched the progress of the Working Groups. Together with the Working Group Leaders, it oversaw the final stages of the study. The Writing Committee submitted its draft final report to the Organizing Committee members and Working Group leaders, who could then ensure that this report appropriately reflects the views of the study participants, and who bear final responsibility for the report's contents. 

Further information on the study and links to the Working Group web pages may be found at {\bf http://www.interactions.org/neutrinostudy}.

\subsection{Charge of the Study}
{\it 
\par
The APS Divisions of Particles and Fields and of Nuclear Physics, together with the APS Divisions of Astrophysics and the Physics of Beams, is organizing a year-long Study on the Physics of Neutrinos, beginning in the fall of 2003.  The Study is in response to the remarkable recent series of discoveries in neutrino physics and to the wealth of experimental opportunities on the horizon.  It will build on the extensive work done in this area in preparation for the 2002 long range plans developed by NSAC and HEPAP, as well as more recent activities, by identifying the key scientific questions driving the field and analyzing the most promising experimental approaches to answering them.  The results of the Study will inform efforts to create a scientific roadmap for neutrino physics. 
\par
The Study is being carried out by four APS Divisions because neutrino physics 
is inherently interdisciplinary in nature.  The Study will consider the 
field in all its richness and diversity.  It will examine 
physics issues, such as neutrino mass and mixing, the number and types of neutrinos, their unique assets as probes of hadron structure, and their roles in astrophysics and cosmology.  It will also study a series of experimental approaches, including long and short baseline accelerator experiments, reactor experiments, nuclear beta-decay and double beta-decay experiments, as well as cosmic rays and cosmological and astrophysical observations.  In addition, the study will explore theoretical connections between the neutrino sector and physics in extra dimensions or at much higher scales. 
\par
The Study will be led by an Organizing Committee and carried out by 
Working Groups.  The Organizing Committee will function as an 
interdisciplinary team, reporting to the four Divisions, with significant 
international participation.  The Study will be inclusive, with all 
interested parties and collaborations welcome to participate. The final 
product of the Study will be a book (or e-book) containing reports from 
each Working Group, as well as contributed papers by the Working Group 
participants.  The Organizing Committee and Working Group leaders will 
integrate the findings of the Working Groups into a coherent summary 
statement about the future.  The Working Groups will meet as necessary, 
with a goal of producing the final report by August 2004. 
\par
The overarching purpose of the Study is for a diverse community of scientists to examine the broad sweep of neutrino physics, and if possible, to move toward agreement on the next steps toward answering the questions that drive the field.  The Study will lay scientific groundwork for the choices that must be made during the next few years. 
}

\begin{table}[p]

{\bf Organizing Committee:} 

\begin{tabular}{|lll|}\hline
Stuart J. Freedman & University of California, Berkeley & Co-chair \\
Boris  Kayser & Fermilab & Co-chair \\
Janet Conrad & Columbia University & \\
Guido Drexlin & University of Karlsruhe &  \\
Belen Gavela &  University Autonoma de Madrid&  \\
Takaaki Kajita & University of Tokyo & \\
Paul Langacker & University of Pennsylvania & \\
Keith Olive &  University of Minnesota &  \\
Bob Palmer &   Brookhaven National Lab &  \\
Georg Raffelt & Max Planck Institute for Physics   & \\
Hamish Robertson & University of Washington &  \\
Stan Wojcicki & Stanford University & \\
Lincoln Wolfenstein & Carnegie-Mellon University &  \\ \hline
\end{tabular}

\vspace*{12pt}
{\bf Working Groups and Group Leaders: }

\begin{tabular}{|ll|}
\multicolumn{2}{l}{\bf Solar and atmospheric neutrino experiments:} \\ 
\hline
   John N. Bahcall & Institute for Advanced Study \\
Joshua R. Klein & University of Texas, Austin \\
\hline\multicolumn{2}{l}{\bf Reactor neutrino experiments:} \\     
\hline
Gabriela Barenboim & University of Valencia \\
Ed Blucher& University of Chicago \\
\hline\multicolumn{2}{l}{\bf Superbeam experiments and development:} \\    
\hline
William Marciano & Brookhaven National Laboratory \\
 Douglas Michael & California Institute of Technology \\
 \hline\multicolumn{2}{l}{\bf Neutrino factory and beta beam experiments and 
development:} \\     
\hline
Stephen Geer & Fermilab \\
  Michael Zisman & Lawrence Berkeley National Laboratory \\
 \hline
\multicolumn{2}{l}{\bf Neutrinoless double beta decay and direct searches 
for neutrino mass:} \\     
\hline
Steven R. Elliott & Los Alamos National Laboratory \\
   Petr Vogel & California Institute of Technology \\
   \hline\multicolumn{2}{l}{\bf Neutrino Astrophysics and Cosmology:} \\    
\hline
Steven Barwick & University of California, Irvine \\
   John Beacom & Ohio State University \\
\hline
\multicolumn{2}{l}{\bf Theory:} \\
\hline
   Rabi Mohapatra  & University of Maryland  \\ \hline
\end{tabular}

\vspace*{12pt}
{\bf Writing Committee}

\begin{tabular}{|lll|} \hline
Janet Conrad & Columbia University & \\
Steve Elliott & Los Alamos National Laboratory & \\
Stuart J. Freedman & University of California, Berkeley & \\
Maury Goodman & Argonne National Laboratory & \\
Andr\'{e} de Gouv\^{e}a & Northwestern University &  \\
Boris  Kayser & Fermilab &  \\
Joshua R. Klein & University of Texas, Austin & \\
Douglas Michael & California Institute of Technology & \\
Hamish Robertson & University of Washington & Chair \\ \hline
\end{tabular}
\caption{Committees}
\label{tab:committees}
\end{table}

\subsection{Sponsors for domestic neutrino science}
\begin{itemize}
\item[$\odot$] \underline{Department of Energy Office of High Energy Physics}
The mission of the High Energy Physics (HEP)
program is to explore the fundamental nature of matter,
energy, space, and time.

\item[$\odot$] \underline{Department of Energy Office of Nuclear Physics}
The DOE Nuclear Physics (NP) program aims to
understand the composition, structure, and properties
of atomic nuclei, the processes of nuclear astrophysics
and the nature of the cosmos. 
\item[$\odot$] \underline{Department of Energy, National Nuclear Security
Administration}
\item[$\odot$]  \underline{National Science Foundation}
\item[$\odot$]  \underline{National Aeronautics and Space Administration}
\end{itemize}

\subsection{Context: Related Studies and Reports} 

\begin{itemize}
\item[$<>$]
The Nuclear Science Advisory Committee's long-range plan, 
``Opportunities in Nuclear Science: A Long-Range Plan for the 
Next Decade.'' \newline {\bf www.sc.doe.gov/henp/np/nsac/nsac.html}

\item[$<>$]
The High-Energy Physics Advisory Panel subpanel report on long-range planning, ``The Science Ahead: The Way to Discovery,'' lays out a roadmap for the U.S. particle physics program over
the next 20 years, also known as the ``Bagger-Barish'' report.
\newline {\bf doe-hep.hep.net/lrp\_panel/index.html}

\item[$<>$]
The DOE ``Office of Science Strategic Plan'' and the 20-year facilities 
roadmap, ``Facilities for the Future of Science: A Twenty-Year Outlook.''   
\newline {\bf www.sc.doe.gov/Sub/Mission/Mission\_Strategic.htm}

\item[$<>$]
The National Research Council (NRC) laid out 11 key scientific questions at the intersection of physics and astronomy in
a report entitled 
``Connecting Quarks with the Cosmos: Eleven Science Questions for the 
New Century.'' 
\newline {\bf www.nationalacademies.org/bpa/projects/cpu/report.}

\item[$<>$]
The OSTP report entitled ``The Physics of the Universe: A Strategic
Plan for Federal Research at the Intersection of Physics and Astronomy''    
is the response of the White House to the NRC Report 
``Connecting Quarks with the Cosmos.''  
One of its recommendations is that NSF and DOE should collaborate to 
``identify a core suite of physics experiments'' for research into 
Dark Matter, neutrinos, and proton decay; and that NSF should take the 
lead on conceptual development and formulation of a scientific roadmap 
for an underground laboratory facility.  
\newline {\bf www.ostp.gov/html/physicsoftheuniverse2}

\item[$<>$]
A National Research Council Report, ``Neutrinos and Beyond: New Windows on 
Nature,'' addresses the scientific motivation for the Ice Cube 
project at the South Pole and for a multipurpose national underground 
laboratory. 
\newline {\bf books.nap.edu/catalog/10583.html}

\item[$<>$]
A HEPAP Subpanel Report, ``Quantum Universe: The Revolution in 21st Century 
Particle Physics'' identifies nine questions for particle physics.
\newline {\bf interactions.org/cms/?pid=1012346.}

\item[$<>$]
A White Paper Report on Using Reactors to search for a value of $\theta_{13}$.
\newline {\bf www.hep.anl.gov/minos/reactor13/reactor13 }

\item[$<>$]
A Fermilab Report, ``The Coming Revolution in Particle Physics "
 Report of the Fermilab Long Range Planning Committee.
\newline
{\bf www.fnal.gov/pub/today/directors\_corner/lrpreportfinal}
\end{itemize}

\begin{twocolumn}
\subsection{Glossary of acronyms}
\begin{itemize}
\item AGS - Alternating Gradient Synchrotron, accelerator at Brookhaven
\item AMANDA - Antarctic Muon And Neutrino Detector Array
\item ANITA - ANtarctic Impulse Transient Antenna 
\item ANTARES - Astronomy with a Neutrino Telescope and Abyss environmental RESearch
\item APS - American Physical Society 
\item BBN - Big Bang Nucleosynthesis 
\item BOONE - BOOster Neutrino Experiment 
\item CC - Charged Current neutrino event 
\item CDF - Collider Detector Facility
\item CERN - European Laboratory for Particle Physics
\item CKM - Cabbibo-Kobayashi-Maskawa 3x3 mixing matrix
\item CMB - Cosmic Microwave Background
\item CHORUS - C(ERN) Hybrid Oscillation Research apparatUS
\item CNGS - C(ERN) Neutrinos to Gran Sasso
\item CPT - Charge conjugation - Parity - Time reversal invariance
\item CUORE - Cryogenic Underground Observatory for Rare Events 
\item D0 - (D-zero) collider experiment at Fermilab intersection region D0
\item DOE - Department of Energy
\item EXO - Enriched Xenon beta-beta decay Observatory
\item FNAL - Fermi National Accelerator Lab
\item GALLEX - GALLium EXperiment  
\item GENIUS - GErmanium liquid NItrogen Underground Study
\item GNO - Germanium Neutrino Observatory 
\item GUT - Grand Unified Theory 
\item GZK  - Greisen Zatsepin Kuzmin cutoff in cosmic ray energy spectrum
\item HELLAZ - HElium at Liquid AZzote temperature
\item HEPAP - High Energy Physics Advisory Panel 
\item ICARUS - Imaging Cosmic and Rare Underground Signals 
\item INO - Indian Neutrino Observatory (proposal)
\item JPARC - Japanese PArticle Research Center
\item K2K - KEK to Super-Kamiokande 
\item KamLAND - Kamioka Liquid scintillator Anti-Neutrino Detector
\item KASKA - Kashiwazaki-Kariwa Reactor Neutrino (proposal) 
\item KATRIN - KArlsruhe TRItium Neutrino Experiment
\item LENS - Low Energy Neutrino Spectroscopy 
\item LEP - Large Electron Proton collider
\item LMA - Large Mixing Angle Solution of the Solar neutrino problem
\item LSND - Liquid Scintillator Neutrino Detector 
\item MINERvA - Main INjector ExpeRiment (neutrino)-A 
\item MINOS - Main Injector Neutrino Oscillation Search 
\item MOON - MOlybdenum Observatory for Neutrinos 
\item MNSP - Maki Nakagawa Sakata Pontecorvo 3x3 mixing matrix
\item MSW - Mikheyev-Smirnov-Wolfenstein  matter-enhancement effect for neutrino oscillations
\item MWE -Meters of Water Equivalent
\item NC - Neutral Current neutrino event
\item NEMO - Neutrino Ettore Majorana Observatory
\item NOvA - NuMI Off-axis (neutrino) Appearance
\item NOMAD - Neutrino Oscillation MAgnetic Detector (CERN) 
\item NSF - National Science Foundation
\item NuMI - Neutrinos at the Main Injector 
\item NuTeV - Neutrinos at the TeVatron 
\item OMB - Office of Management and Budget
\item OPERA - Oscillation Project with Emulsion-tRacking Apparatus 
\item OSTP - Office of Science and Technology Policy
\item QCD - Quantum ChromoDynamics
\item P5 - Particle Physics Project Prioritization Panel
\item QE - Quasi-Elastic neutrino event
\item R\&D  - Research and Development
\item RICE - Radio Ice Cerenkov Experiment 
\item SAGE - (Soviet) russian American Gallium Experiment 
\item SAGENAP - Scientific Assessment Group for Experimental Non-Accelerator Physics 
\item SLC - Stanford Linear Collider
\item SM - Standard Model of  particles and fields
\item SN(e) - Supernova(e) 
\item SNO - Sudbury Neutrino Observatory
\item SPS - CERN Super Proton Synchrotron
\item SSM - Standard Solar Model
\item Super-Kamiokande - Super-Kamioka Nucleon Decay Experiment
\item SUSY - SUper SYmmetry
\item T2K - Tokai to Kamioka long-baseline experiment at JPARC 
\item WMAP - Wilkinson Microwave Anisotropy Probe 
\end{itemize}
\end{twocolumn}

\end{onecolumn}
\end{document}